\documentclass[
aps,
nofootinbib,
prd,
superscriptaddress,
tightenlines,
notitlepage,
twocolumn,
showpacs,
floatfix
]{revtex4-1}

\usepackage[colorlinks]{hyperref}
\usepackage{tabularx}
\usepackage{graphicx}
\usepackage{subfigure}
\usepackage{float}
\usepackage{amssymb,amsmath,bm,tensor,braket}
\usepackage{xcolor}
\usepackage[varg]{txfonts}
\usepackage{enumerate}
\usepackage{mathtools}
\usepackage[capitalize]{cleveref}
\usepackage[utf8]{inputenc}
\usepackage[normalem]{ulem}
\usepackage{esvect}
\usepackage{xcolor}
\usepackage{soul}
\usepackage{dcolumn}% Align table columns on decimal point
\usepackage[T1]{fontenc}

\makeatother
\begin{document}
% \preprint{APS/123-QED}

% \title{Efficient Data Products for Gravitational-Wave Posteriors via Neural Density Estimation}
% \title{Towards explainable and efficient neural density estimator for gravitational waves}
\title{Lightweight posterior construction for gravitational-wave catalogs \\
with the Kolmogorov-Arnold network}

\author{Wenshuai Liu}
\affiliation{Department of Astronomy, School of Physics, Peking University, Beijing 100871, China}%

\author{Yiming Dong}\email[]{ydong@pku.edu.cn}
\affiliation{Department of Astronomy, School of Physics, Peking University, Beijing 100871, China}%
\affiliation{Kavli Institute for Astronomy and Astrophysics, Peking University, Beijing 100871, China}

\author{Ziming Wang}
\affiliation{Department of Astronomy, School of Physics, Peking University, Beijing 100871, China}%
\affiliation{Kavli Institute for Astronomy and Astrophysics, Peking University, Beijing 100871, China}

\author{Lijing Shao}\email[]{lshao@pku.edu.cn}
\affiliation{Kavli Institute for Astronomy and Astrophysics, Peking University, Beijing 100871, China}%
\affiliation{National Astronomical Observatories, Chinese Academy of Sciences, Beijing 100012, China}

%\date{\today}
% It is always \today, today,
% but any date may be explicitly specified

%---------------------------------------------------------------------
\begin{abstract}
Neural density estimation has seen widespread applications in the
gravitational-wave (GW) data analysis, which enables real-time parameter
estimation for compact binary coalescences and enhances rapid inference for
subsequent analysis such as population inference.  In this work, we explore the
application of using the Kolmogorov-Arnold network (KAN) to construct efficient
and interpretable neural density estimators for lightweight posterior
construction of GW catalogs.  By replacing conventional activation functions
with learnable splines, KAN achieves superior interpretability, higher accuracy,
and greater parameter efficiency on related scientific tasks.  Leveraging this
feature, we propose a KAN-based neural density estimator, which ingests
megabyte-scale GW posterior samples and compresses them into model weights of
tens of kilobytes.  Subsequently, analytic expressions requiring only
several kilobytes can be further distilled from these neural network weights
with minimal accuracy trade-off. In practice, GW posterior samples with fidelity
can be regenerated rapidly using the model weights or analytic expressions for
subsequent analysis.  Our lightweight posterior construction strategy is
expected to facilitate user-level data storage and transmission, paving a path
for efficient analysis of numerous GW events in the next-generation GW
detectors.
\end{abstract}
%---------------------------------------------------------------------

\maketitle

%---------------------------------------------------------------------
\section{Introduction}
\label{Sec1_Introduction}
%---------------------------------------------------------------------

%==============introduction of GW

Since the first detection of the gravitational-wave (GW) signal from a binary
black hole (BBH) merger~\cite{LIGOScientific:2016aoc}, GWs have opened a new era
in our exploration of the Universe.  The ground-based GW detectors, 
LIGO~\cite{LIGOScientific:2014pky}, Virgo~\cite{VIRGO:2014yos}, and
KAGRA~\cite{KAGRA:2018plz}, have completed three observing
runs~\cite{LIGOScientific:2018mvr, LIGOScientific:2020ibl,
LIGOScientific:2021usb, KAGRA:2021vkt}, and together with the first part of the
fourth run~\cite{LIGOScientific:2025hdt, LIGOScientific:2025slb}, more than 200
GW events from compact binary coalescences (CBCs) are reported.  These
observations have driven transformative advances across multiple domains of
astrophysics~\cite{LIGOScientific:2018jsj, LIGOScientific:2020kqk,
KAGRA:2021duu}, cosmology~\cite{ligo2017gravitational, DES:2019ccw,
LIGOScientific:2019zcs, LIGOScientific:2021aug}, and fundamental
physics~\cite{Yagi:2013awa, LIGOScientific:2018cki, LIGOScientific:2021sio}.

%==============Data analysis

Behind these groundbreaking discoveries lie the advancements in instruments, as
well as data analysis techniques.  Bayesian inference constitutes the
fundamental framework in GW data analysis, providing a principled and systematic
approach to parameter estimation and population
inference~\cite{thrane2019introduction}.  In Bayesian inference, the parameter
estimation is calculating the conditional distribution, the so-called posterior
$P(\boldsymbol{\theta}|d)$, of the parameters $\boldsymbol{\theta}$ given the
observed data $d$.  By employing Bayes' theorem, the posterior can be obtained
given the prior $\pi(\boldsymbol{\theta})$ and the likelihood function
$\mathcal{L}(d|\boldsymbol{\theta})$, via $P(\boldsymbol{\theta}|d) =
{\mathcal{L}(d|\boldsymbol{\theta})\pi(\boldsymbol{\theta})}/{\mathcal{Z}(d)}$,
where $\mathcal{Z}(d)$ is the normalization factor known as the evidence.
Usually, it is unfeasible to directly calculate the exact expression of the
posterior. Luckily, with advanced stochastic sampling methods such as Markov
Chain Monte Carlo (MCMC)~\cite{Christensen:1998gf,Sharma:2017wfu} and nested
sampling~\cite{Skilling:2004pqw,Skilling:2006gxv}, we can draw posterior samples
from this distribution. Posterior samples constitute key data products in GW
catalogs underpinning subsequent scientific investigation in GW astronomy.

%==============catalog and HBA

As GW detections become routine, the focus of GW astrophysics is expanding from
analyzing individual events to conducting systematic studies on the entire
population of sources.  This shift necessitates a compilation of comprehensive
and accessible GW catalogs.  Growing catalogs call for powerful statistical
techniques, like hierarchical Bayesian analyses~\cite{Mandel:2018mve,
thrane2019introduction,Vitale:2020aaz,Wang:2024xon,Gerosa:2024isl},
to unravel the collective properties of compact binaries.  Distributions of compact binary source
parameters, such as masses~\cite{KAGRA:2021duu,Golomb:2021tll,Dong:2024zzl} and
spins~\cite{KAGRA:2021duu,Zhu:2017znf,Tong:2022iws} provide critical constraints
on models of stellar evolution and compact object formation.  Furthermore,
combining information from multiple events allows for stringent tests of
Einstein's theory of general relativity~\cite{LIGOScientific:2021sio} and
precise constraints on the properties of dense matter, such as the equation of
state of neutron stars~\cite{Landry:2018prl, Wysocki:2020myz, Shao:2022koz,
Wang:2024xon}.  Therefore, efficiently constructing and utilizing GW catalogs is
crucial to advance our understanding of the Universe, especially in the era of
next-generation GW detectors when many more events are detected.

%==============dificulties in the future

Next-generation ground-based GW detectors, such as the Einstein Telescope
(ET)~\cite{Punturo:2010zz, Branchesi:2023mws, Abac:2025saz} and Cosmic Explorer
(CE)~\cite{Reitze:2019iox, Evans:2021gyd}, will achieve unprecedented
sensitivities, unlocking new scientific
opportunities~\cite{Sathyaprakash:2019yqt, Kalogera:2021bya}.  They will boost
the detection rates for binary neutron-star and black-hole mergers, reaching
$10^{5}$--$10^{6}$ events per year~\cite{Himemoto:2021ukb,Samajdar:2021egv,
Hu:2022bji,Borhanian:2022czq,Pieroni:2022bbh,Ronchini:2022gwk,Iacovelli:2022bbs,
Begnoni:2025oyd}. Considering that standard offline Bayesian inference for a single
event already demands high computational resources, the high event rates
expected in the next-generation GW detectors will introduce new computational
challenge for data analysis. As predicted, performing standard Bayesian
inference on a one-month catalog requires $10^{13}$--$10^{15}$ CPU hours,
imposing a substantial burden on computing infrastructure, electricity cost, and
environmental impact~\cite{Hu:2024mvn}.  Efficient, accelerated Bayesian methods
are therefore urgently needed.  Besides, from the perspective of data storage
and transmission, posterior-sample-based GW catalogs with millions of events
could easily reach a terabyte-scale volume, placing significant burdens on users
for both transmission and storage.

%==============machine learning

The development of machine learning techniques offers a possible solution to the
data challenges posed by next-generation GW detectors. Machine learning has
already demonstrated its great potential in GW detection~\cite{George:2017pmj,
Huerta:2017kez, Xia:2020vem, Ma:2023ctz, Ma:2024log, Wang:2025xvj}, parameter
estimation~\cite{Gabbard:2019rde, Green:2020hst, DEmilio:2021laf,
Santoliquido:2025lot, Green:2020dnx, Dax:2021tsq, Dax:2024mcn}, and subsequent
astrophysical researches~\cite{Leyde:2023iof, mcewen2021machine, Talbot:2020oeu,
Mould:2022ccw, Ventagli:2024xsh, Tang:2024orq, Hu:2025vlp}.  Specifically,
neural posterior estimation has been widely adopted across multiple GW Bayesian
inference tasks~\cite{Dax:2021tsq, Dax:2024mcn, Leyde:2023iof,
mcewen2021machine}, accelerating the analysis by orders of magnitude.  In
particular, neural posterior estimation with normalizing
flows~\cite{Green:2020dnx, Dax:2021tsq, Dax:2024mcn, Leyde:2023iof,
Langendorff:2022fzq, kobyzev2020normalizing} achieves real-time Bayesian
inference for binary black hole mergers and binary neutron star mergers, opening
new avenues for multi-messenger astronomy. In the era of next-generation
detectors, neural posterior estimation has also been explored to address data
challenges such as overlapping signals~\cite{Langendorff:2022fzq, Hu:2025vlp}
and long-duration signal analysis~\cite{Hu:2024lrj}.  For neural posterior
estimation, it needs a neural-network-based density estimator to represent
distributions. These neural density estimators are usually built on
generative-model architectures, such as variational
autoencoders~\cite{Kingma:2013hel, Gabbard:2019rde}, normalizing flows, and flow
matching frameworks~\cite{Chen:2018wjc, lipman2023flow, Liang:2024new}. For
instance, normalizing flows learn a sequence of invertible transformations that
map a simple base distribution to the complex target distribution, enabling both
fast sampling and exact density evaluation.  By efficiently capturing
high-dimensional, multimodal distributions, neural density estimators show great
promise for alleviating the computational bottlenecks in GW Bayesian inference.

%==============our method

In this work, we propose a Kolmogorov-Arnold network
(KAN)~\cite{liu2025kan,liu2024kan2} based neural density estimator and explore
its potential for constructing high-fidelity, low-storage surrogate GW catalogs.
KAN is a machine-learning architecture inspired by the Kolmogorov-Arnold
representation theorem~\cite{arnold2009representation,braun2009constructive}.
Rather than using fixed activations in conventional multi-layer perceptrons
(MLPs), KAN employs edge-wise learnable activation functions.  This modification
enables KAN to achieve superior interpretability, higher accuracy, and greater
parameter efficiency compared to conventional architectures on small-scale
artificial intelligence (AI) for scientific tasks. Furthermore, this design also
allows the network to converge toward interpretable analytic expressions, a
process referred to as the symbolification of the neural network.  Currently,
KANs are used in fields such as fluid dynamics~\cite{Toscano2025, Wang_2025},
nuclear physics~\cite{PhysRevC.111.024316}, complex dynamical
systems~\cite{PhysRevResearch.7.023037}, molecular property prediction~\cite{li2025KAGNN},
and cosmology~\cite{Cui:2025rri, Jones:2025obf}, and
KAN-based machine learning architectures are also rapidly
advancing~\cite{yang2025kolmogorovarnold, fang2025kaa, huang2025timekan}.  The
rapid development of KAN offers a promising approach to enhance the
interpretability of neural networks in AI for science tasks.  In our case, to
support downstream studies based on GW posteriors, one can download compact,
encoded files from our surrogate GW catalogs and rapidly resample the posterior
distributions locally.  This strategy enables efficient catalog construction
while dramatically reducing the demands on data transmission and storage.

Leveraging KAN's efficient representation and unique symbolification capability,
our KAN-based neural density estimator can provide two compressed
representations of GW posterior samples.  The first is the set of {\it neural
network weights}, which prioritize high fidelity to support precision-critical
analysis tasks.  The second representation, {\it analytic fits}
of the posterior distributions, maximizes compression by encoding the posteriors
in compact, closed-form analytic expressions, though with a small trade-off in accuracy.
Both products enable fast posterior resampling, and their reliability is validated
in downstream analysis tasks through GW population inference, exhibiting great
potential in efficient and robust GW catalog construction for future GW observations.
%==============Structure of the paper

The paper is organized as follows.  In Sec.~\ref{Sec2_Method}, we introduce the
machine learning techniques underlying the KAN-based neural density estimator
and present the flowchart for efficient GW catalog construction.  We showcase
three verifiable simple examples in Sec.~\ref{Sec3_Experiments} to demonstrate
the feasibility of the proposed method, validating the effectiveness and
accuracy of analytic probability density function (PDF) fitting using the KAN
network.  In Sec.~\ref{Sec4_Gravitational_Wave}, we present the compression
results based on the KAN-based neural density estimator on posterior samples
from some current GW events and further validate the reliability of the
compact data products by performing population inference tasks using the
resampled samples from the compressed data products.  Finally,
Sec.~\ref{Sec5_Summary} summarizes our findings and discusses potential
directions for future improvements.

%=========================================================
\begin{figure*}
    \centering
    \includegraphics[width=\linewidth]{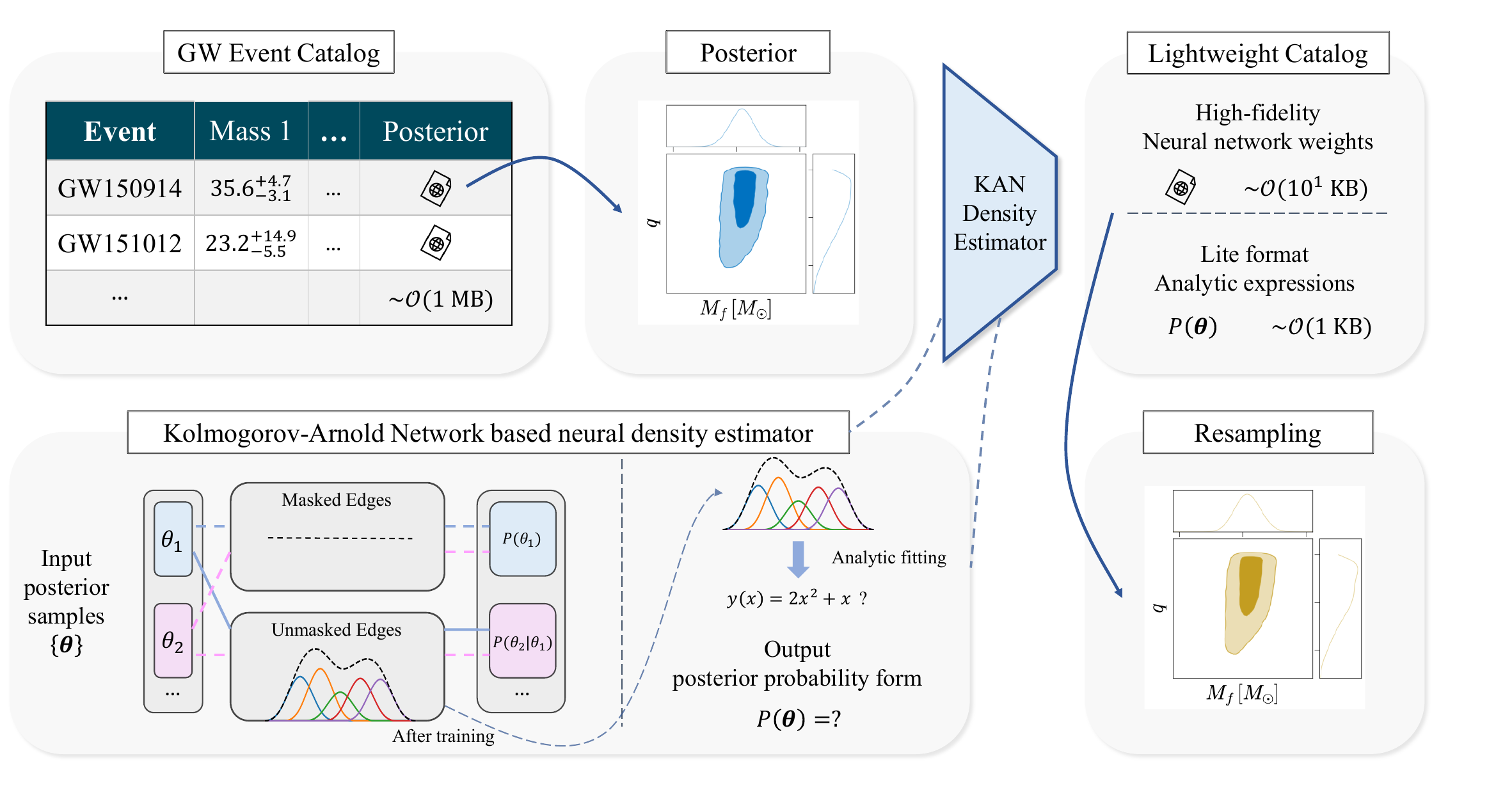}
    \caption{{\bf Flowchart for lightweight GW catalog construction.}  Starting
    from raw posterior samples of a single GW event as our training set, we
    first train the KAN-based neural density estimator to minimize the average
    negative log-posterior.  As shown in the box at the bottom left, we replace
    MADE's unmasked edges with KAN's learnable nonlinear edges (a linear
    combination of B-spline curves).  Once the loss has converged, the analytic
    expression of each unmasked edge is  obtained and  synthesized through the
    network topology to get analytic expressions of all output nodes:
    $\ln(a_{ij})$, $\mu_{ij}$, and $-2\ln(\sigma_{ij})$; see main text for
    details.  Expressions of all output nodes are combined to obtain the
    analytic joint PDF.  After training and symbolification, we store the neural
    network weights, typically of size $\mathcal{O}\,(10^{1} \, {\rm KB})$, and
    analytic expressions of output nodes, typically of size $\mathcal{O}\,(1 \,
    {\rm KB})$, as compact data products.  Users can download these data
    products and use them to compute probability density or resample posterior
    samples for downstream analysis tasks.}
    \label{fig:1}
\end{figure*}
%=========================================================

%---------------------------------------------------------------------
\section{KAN-based neural density estimator}
\label{Sec2_Method}
%---------------------------------------------------------------------

In this section, we introduce the method to compress the posterior samples
through a KAN-based neural density estimator.  First, we introduce
KAN~\cite{liu2025kan,liu2024kan2} in Sec.~\ref{Sec2.1_KAN}.  We explain how KAN
improves upon conventional neural networks and enables symbolification, after
which the entire model is represented merely by the analytic expressions of its
output nodes.  In Sec.~\ref{Sec2.2_MADE}, we present the Masked Autoencoder for
Distribution Estimation (MADE)~\cite{germain2015made}, one of the core building
blocks commonly employed in normalizing flows~\cite{Papamakarios:2017tec}.  MADE
serves as our base density estimator architecture, which we then enhance using
the KAN framework.  Section~\ref{Sec2.3_Resampling} describes how to perform
rapid resampling using the neural network weights and/or the analytic
expressions.  We outline the workflow for lightweight GW catalog construction in
Fig.~\ref{fig:1}.  
% Starting from raw posterior samples of a single GW event as our training set, we
% first train the KAN-based neural density estimator.  Once the loss has
% converged, we extract the analytic representation of the trained model.  We
% store both the network weights and the analytic expressions of the output nodes
% as compact data products.  Thanks to the parameter efficiency of KAN, the
% outcome requires much less storage than the original posterior samples.  Users
% can download these data products and use them to compute probability density or
% resample posterior samples for downstream analysis tasks.

%==============KAN

%---------------------------------------------------------------------
\subsection{Kolmogorov-Arnold Network}
\label{Sec2.1_KAN}
%---------------------------------------------------------------------

KAN is a novel machine-learning architecture inspired by the Kolmogorov-Arnold
representation theorem~\cite{arnold2009representation, braun2009constructive}. 
Unlike conventional MLPs that rely on fixed activation functions at each neuron,
KAN introduces learnable activation functions on the edges connecting neurons
between layers.  These edge-wise functions are parameterized as a linear
combination of B-spline curves~\cite{gordon1974b}, with coefficients being
learnable parameters that are optimized during training.  This design allows KAN
to capture complex nonlinear relationships between inputs and outputs more
flexibly, leading to enhanced accuracy and greater parameter efficiency,
particularly in small-scale scientific AI applications.

A key feature of KAN is its capability for network symbolification, a two-stage
process involving function matching and synthesis.  First, the trained B-spline
activation functions on each edge are matched to a predefined library of
elementary mathematical functions (e.g., polynomial, trigonometric, and
exponential functions) to derive their analytic representations.  Subsequently,
these individual analytic expressions are composed according to the network's
topology, yielding closed-form analytic expressions that describe the network's
outputs as explicit functions of its inputs~\cite{liu2025kan}.
This process significantly enhances the interpretability of the neural network,
making its internal mechanism more transparent and analyzable.  Leveraging KAN's
efficient function representation and unique symbolification capabilities, we
propose a KAN-based neural density estimator.  Our objective is to harness these
advantages to achieve substantial storage compression of GW posterior samples,
thereby offering a new approach for constructing lightweight and user-friendly
GW catalogs.

%==============MADE

%---------------------------------------------------------------------
\subsection{Masked Autoencoder for Distribution Estimation}
\label{Sec2.2_MADE}
%---------------------------------------------------------------------

To model the probability distribution of the posterior samples, we employ the
MADE framework.  MADE facilitates density estimation by decomposing the joint
probability distribution autoregressively~\cite{germain2015made}. For
a $D$-dimensional parameter vector $\boldsymbol{\theta}$, this decomposition is
expressed as 
%--
\begin{equation}
    P(\boldsymbol{\theta}) = \prod_{i=1}^D P_{i}(\theta_i |
    \boldsymbol{\theta}_{<i}) \,, \label{eq:2}
\end{equation}
%--
where $\boldsymbol{\theta}_{<i}$ denotes the set of parameters $\{ \theta_1,
\theta_2, \cdots, \theta_{i-1} \}$. MADE approximates each conditional
probability $P_{i}$ with a Gaussian mixture model (GMM)~\cite{mcewen2021machine,germain2015made,papa2016fast},
such that
%--
\begin{equation}
    P_{i}(\theta_i | \boldsymbol{\theta}_{<i}) \approx \sum_{j=1}^G
    a_{ij}(\boldsymbol{\theta}_{<i}) \mathcal{N}\left(\theta_i \,\big|\,
    \mu_{ij}(\boldsymbol{\theta}_{<i}),
    \sigma_{ij}^2(\boldsymbol{\theta}_{<i})\right) \,.  \label{eq:3}
\end{equation}
%--
Here, $a_{ij}$, $\mu_{ij}$, and $\sigma_{ij}$ are the mixture weights, means,
and standard deviations of the Gaussian components, respectively, all of
which are functions of the preceding parameters $\boldsymbol{\theta}_{<i}$.
The number of Gaussian components used in each GMM is denoted as $G$.

A neural network, denoted by its weights $w$, is trained to output the
parameters of these GMMs.  To ensure positivity of $a_{ij}$ and $\sigma_{ij}$,
the network predicts $\ln(a_{ij})$ and $-2\ln(\sigma_{ij})$ alongside
$\mu_{ij}$.  The complete network-approximated distribution is denoted as
$P_w(\boldsymbol{\theta})$.  The crucial autoregressive property is enforced by
applying binary masks to the network's connections, ensuring that each
conditional probability $P_{i}$ depends only on $\boldsymbol{\theta}_{<i}$. 
Alternative approaches for enforcing autoregressive constraints include masked
convolutions~\cite{van2016pixel} and causal convolutions~\cite{oord2016wavenet}.
If the target distribution is conditional
$P(\boldsymbol{\theta}|\boldsymbol{\phi})$, the conditioning variables
$\boldsymbol{\phi}$ need to be connected to the network through unmasked
connections.

In our implementation, we enhance the standard MADE architecture by replacing
its linear layers with KAN layers, creating a KAN-based MADE.  This allows us to
later extract analytic expressions for the PDF.  Leveraging KAN's parameter
efficiency, our framework is expressive enough even with a light network
structure and also only requires small storage space.  The network is trained by
minimizing the average negative log-posterior,
$-\mathbb{E}_{P(\boldsymbol{\theta})} \big[\ln P_w(\boldsymbol{\theta}) \big]$,
using raw posterior samples as the training set.  After loss convergence, we
obtain two compact data products:
%---
\begin{enumerate}[(i)]
    \item \textbf{Neural Network Weights:} The trained weights $w$ of the
    KAN-based MADE, which provide a high-fidelity representation of the
    posterior distribution.
    \item \textbf{Analytic Expressions:} By applying KAN's symbolification
    process, we derive analytic expressions for all  GMM parameters ($a_{ij}$,
    $\mu_{ij}$, $\sigma_{ij}$) as functions of $\boldsymbol{\theta}_{<i}$.
    These expressions offer a maximally compressed, interpretable representation
    of the posterior.  We also offer functionality to express the joint
    probability distribution $P_w(\boldsymbol{\theta})$ in closed form using
    these saved expressions according to Eq.~(\ref{eq:2}) and Eq.~(\ref{eq:3}),
    in order to facilitate subsequent probability density evaluation.
\end{enumerate}
%---
These data products enable users to reconstruct the posterior distribution for
further analysis with little storage and computational overhead, compared with
the original posterior samples in the GW catalogs.

%==============resampling method

%---------------------------------------------------------------------
\subsection{Resampling Using Neural Network Weights and Analytic Expressions}
\label{Sec2.3_Resampling}
%---------------------------------------------------------------------

Both the neural network weights and the analytic expressions can be used to
generate new posterior samples and evaluate the probability density.  The
resampling process is autoregressive: for each dimension $i=1, \dots, D$,
$\theta_i$ is sampled from the conditional distribution $P_i(\theta_i |
\boldsymbol{\theta}_{<i})$ using the already-sampled values of
$\boldsymbol{\theta}_{<i}$.  The resampling process using analytic expressions
is more efficient than resampling with the neural network weights.

%non-analytic

For the resampling with neural network weights, the sample generation process is
accomplished iteratively according to Eq.~(\ref{eq:2}) where
$\boldsymbol{\theta}$ is set to $\boldsymbol{0}$ in the beginning.  During the
$i$-th iteration, $\boldsymbol{\theta}_{<i}$ is already generated and
$\boldsymbol{\theta}_{\geq i}$ is still $\boldsymbol{0}$.  Then
$\{\boldsymbol{\theta}_{<i},\boldsymbol{0}\}$ is fed into the network to obtain
$a_{i,1\leq j\leq G}$, $\mu_{i,1\leq j\leq G}$, and $\sigma_{i,1\leq j\leq G}$.
Next, $\theta_i$  is sampled from the conditional distribution $P_i(\theta_i |
\boldsymbol{\theta}_{<i})$.  However, in each sequential iteration, all network
outputs are computed, introducing computational inefficiency as many node
computations---those not related to $P_{i}$---are unnecessary.  If we assume
that the network does not contain hidden layers (as we do), then the
computational complexity of resampling is $\mathcal{O}\, \big(GD^3 \big)$.
This is because if the number of nodes in the input layer is $D$, the number of
nodes in the output layer is proportional to $GD$, and the number of iterations
equals $D$.  Probability density computation using a neural network requires one
forward computation to get all the GMM parameters and the complexity is
$\mathcal{O}\, \big(GD^2 \big)$.  Then $P_w(\boldsymbol{\theta})$ is computed
according to Eq.~(\ref{eq:2}) and Eq.~(\ref{eq:3}).

%analytic

For the resampling with analytic expressions, it enables more efficient sampling
through direct parameter computation.  Specifically, in each iteration, the
first $i-1$ dimensions can be directly substituted into analytic expressions of
$a_{i,1\leq j\leq G}$, $\mu_{i,1\leq j\leq G}$, and $\sigma_{i,1\leq j\leq G}$.
This optimized process eliminates computation of output nodes that are not
related to $P_{i}$.  It reduces sampling complexity to $\mathcal{O}\,(GD)$ and
provides substantial efficiency gains in high-dimensional cases.  Probability
density computation using analytic expressions simply requires substituting
observed values into the analytic expression of $P_w(\boldsymbol{\theta})$.
This direct evaluation process requires only basic math operations.
Furthermore, the analytic $P_w(\boldsymbol{\theta})$ provides mathematical
tractability, permitting direct manipulation of the PDF through algebraic
operations.

%---------------------------------------------------------------------
\begin{figure*}[]
    \centering
    \includegraphics[width=\linewidth]{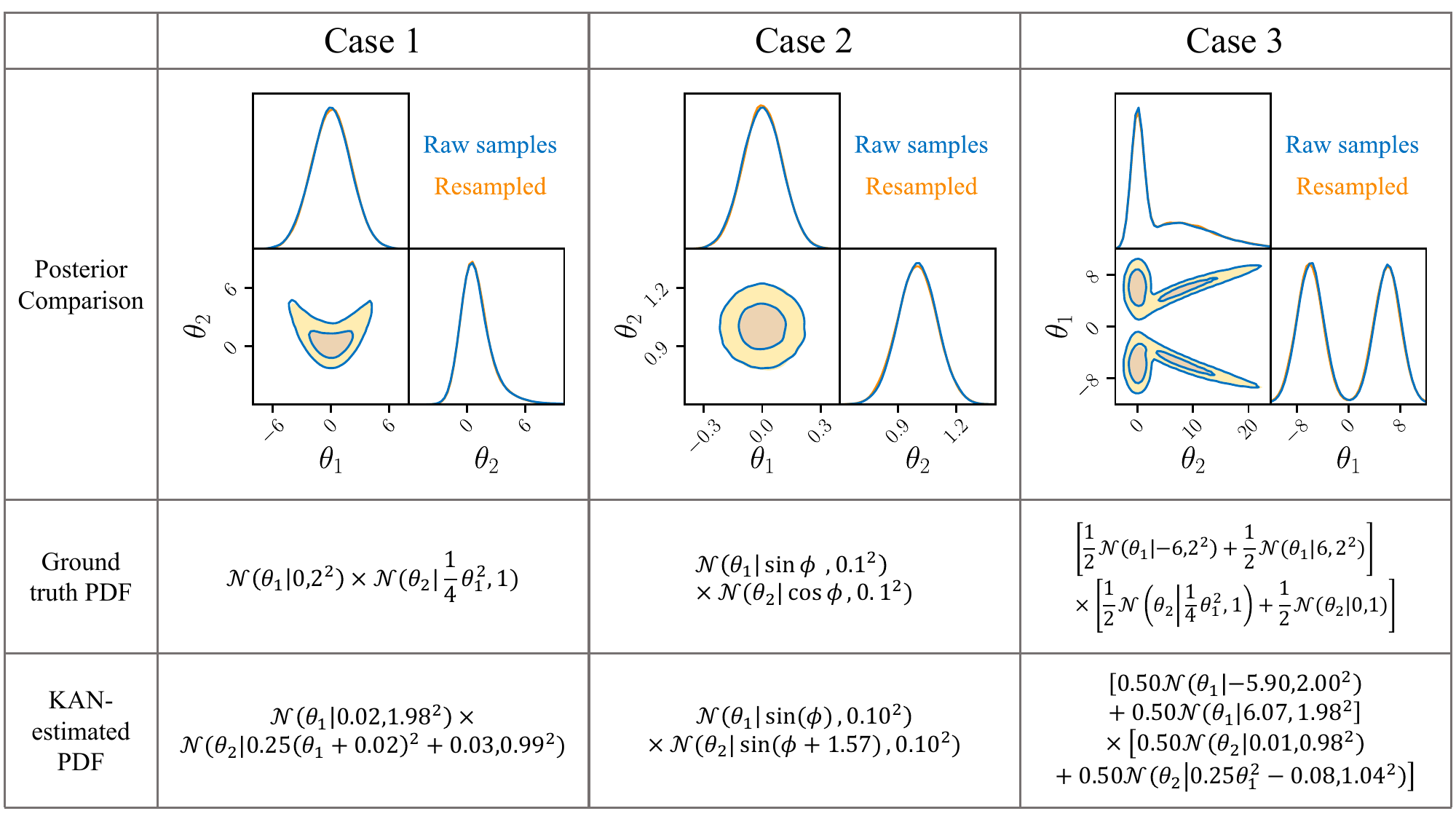}
    \caption{ {\bf Results of three verifiable cases.} Marginalized one- and
    two-dimensional distributions are shown in corner plots, comparing samples
    resampled from fitted expressions (orange) and samples generated from
    ground-truth distributions (blue).  Both of them contain 40,000 samples.  In
    Case 2, $\phi$ is fixed at 0 during the sampling process.  Contour lines in
    the two-dimensional joint distributions delineate the $50\%$ and $90\%$
    credible regions.  The second row displays the ground-truth PDFs of the
    three cases, where $\mathcal{N}(\mu,\sigma^2)$ denotes a Gaussian
    distribution with mean value $\mu$ and variance $\sigma^2$.  The third row
    shows the analytic expressions of the fitted PDFs chosen by our KAN-based
    neural density estimator where the coefficients are rounded to two decimal
    places.}
    \label{fig:2}
\end{figure*}
%---------------------------------------------------------------------

%================================================Experiments

%---------------------------------------------------------------------
\section{Case Tests}
\label{Sec3_Experiments}
%---------------------------------------------------------------------

Before applying the KAN-based neural density estimator to real GW posterior
data, we construct three illustrative cases to test the reliability of our model
and explore its ability to fit posterior distributions into analytic
expressions. We choose a unimodal distribution, a conditional unimodal
distribution, and a bimodal distribution as our test cases~\cite{Papamakarios:2017tec}.
In each example, we start with the ground-truth PDF to obtain the raw samples,
and then use these as input for the KAN-based neural
density estimator to fit for the distribution's analytic form. It is important
to note that we do not specify the analytic form of the distribution in advance.
Instead, we allow the KAN-based neural density estimator to freely explore the
fitting expressions without any additional input information from human. 

The PDFs of the three selected examples are as follows. The first example of a
unimodal distribution has a PDF,
%--
\begin{equation}
    p(\theta_1, \theta_2) = \mathcal{N}(\theta_1 | 0, 2^2) \times \mathcal{N}
    \Big(\theta_2 \,\big|\, \frac{1}{4} \theta_1^2, 1 \Big)\, .  
    \label{eq:4}
\end{equation}
%--
Its distribution forms a crescent shape, and we aim to use this example to test
the method's capability of accommodating non-Gaussian distributions. The PDF of the second
example, a conditional unimodal distribution, reads
%--
\begin{equation}
    p(\theta_1, \theta_2 \,|\, \phi) = \mathcal{N} \big(\theta_1 | \sin \phi,
    0.1^2 \big) \times \mathcal{N} \big(\theta_2 | \cos \phi, 0.1^2 \big)\, .
    \label{eq:5}
\end{equation}
%--
This distribution is conditional on $\phi$. Although an unconditional neural
density estimator is sufficient for this work, exploring conditional density
estimators is more important for future applications in the development of
neural posterior estimation algorithms.  The PDF of the third example, a bimodal
distribution, is
%--
\begin{equation}
    \begin{aligned}
            p(\theta_1, \theta_2)  =& \left[ \frac{1}{2} \mathcal{N}
            \big(\theta_1 | -6, 2^2 \big) + \frac{1}{2} \mathcal{N}\big(\theta_1
            | 6, 2^2 \big) \right] \\
            & \times \left[ \frac{1}{2} \mathcal{N} \Big(\theta_2 \,\big|\,
            \frac{1}{4} \theta_1^2, 1\Big) + \frac{1}{2} \mathcal{N}
            \big(\theta_2 | 0, 1\big) \right] \, .
    \end{aligned}
    \label{eq:6}
\end{equation}
%--
This example involves a highly distorted bimodal distribution, which is used to
test the neural density estimator's fitting ability with multimodal
distributions. Beyond these examples, we also present one additional illustrative example based on an
exponential distribution in Appendix~\ref{AppendixA} to more comprehensively
demonstrate the effectiveness of this method in modeling diverse distributions.

The verification process is nearly identical to the workflow shown in
Fig.~\ref{fig:1}.  We start with raw samples and obtain the encoded data
products.  The difference lies in the fact that in the verifiable examples, we
know the ground-truth PDFs, which allows us to compare them with the analytic
fitting expressions from KAN. In the tested examples, for the KAN-based MADE
network architecture, we include only the input and output layers, without any
intermediate hidden layers.  In all three examples, each unmasked edge of the
network is set to contain eight cubic B-spline functions.  We find that, due to
the powerful fitting ability of the learnable spline edges in KAN, this simplest
MADE design is already remarkable to support high-precision distribution
fitting.

The results are shown in Fig.~\ref{fig:2}, where we present the distribution
expressions fitted by the KAN-based neural density estimator in the last row, 
along with the graphic distribution results obtained by resampling from these
fitted expressions.  As we can see, even without any additional information, KAN
successfully captures accurately the underlying analytic fitting form.  In the
applications to GW catalogs, compared to the raw samples, the encoded analytic
data products offer significant advantages in terms of storage and allow for
more flexibility in determining the number of resampled points to meet varying
computational needs.  In the analytic fitting results, the algorithm uses one
Gaussian component for fitting in the first and second examples, while it uses
two Gaussian components for fitting in the third example.

\begin{figure}[]
    \centering
    \includegraphics[width=0.7\linewidth]{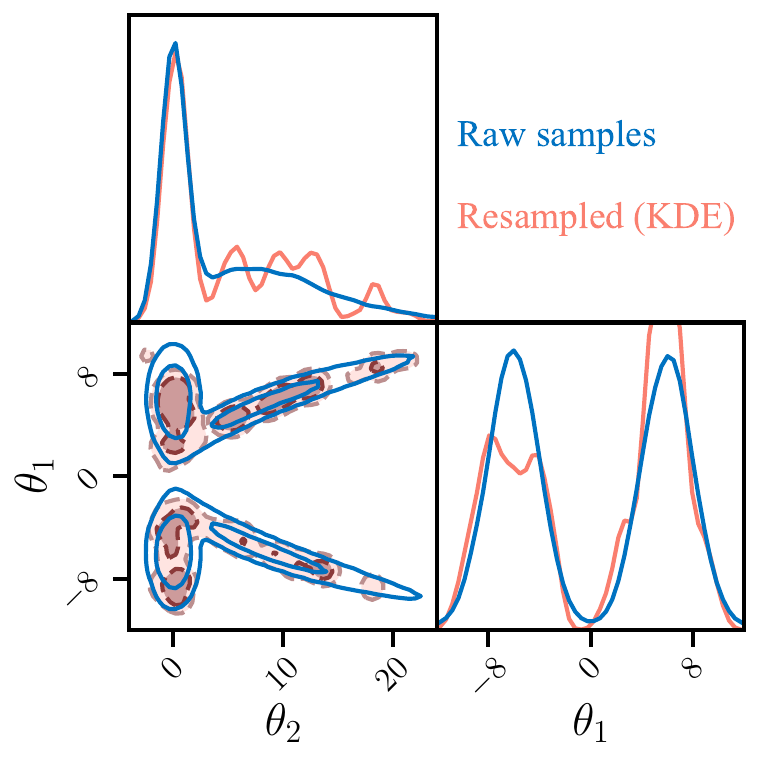}
    \caption{{\bf Results of KDE for Case 3.} The 110 samples are utilized in a
    Gaussian kernel with a bandwidth of 0.35.  The KDE is accomplished using the
    \texttt{scikit-learn} package.  Marginalized one- and two-dimensional
    distributions are shown in the corner plot, comparing samples resampled
    using KDE (red) and samples generated from the ground-true distribution
    (blue).  Both of them contain 40,000 samples.  Contour lines in the
    two-dimensional joint distributions delineate the $50\%$ and $90\%$ credible
    regions. }
    \label{fig:3}
\end{figure}

To further demonstrate the storage efficiency, we conduct a controlled
comparison between storing samples and storing the encoded products using Case
3. Encoded analytic data products exhibit exceptional storage efficiency, 
requiring only $\sim 1\,{\rm KB}$ of storage. The same storage corresponds to
approximately 110 samples.  For comparison with other methods, we present the
distribution results from expanding these 110 samples using kernel density
estimation (KDE) in Fig.~\ref{fig:3}. Clearly, the posterior distribution
represented by these 110 samples exhibits significant fluctuations and spurious
structures, far from accurately capturing the true distribution. Our KAN-based
density estimator (see the third corner plot in Fig.~\ref{fig:2}), in contrast,
offers clear advantages in interpretability and storage efficiency.

%---------------------------------------------------------------------
%---------------------------------------------------------------------

%---------------------------------------------------------------------
\section{Lightweight  posterior construction for GW catalogs}
\label{Sec4_Gravitational_Wave}
%---------------------------------------------------------------------

In this section, we apply the KAN-based neural density estimator to real GW
data.  We select six GW parameters---chirp mass ${\cal M}$, mass ratio $q$, two
spin magnitudes $\chi_1$ and $\chi_2$, and two spin inclinations $\theta_1$ and
$\theta_2$---and attempt to encode the posterior distributions of ten GW events:
GW150914, GW151012, GW151226, GW170104, GW170608, GW170729, GW170809, GW170814,
GW170818, and GW170823~\cite{LIGOScientific:2021usb}.  Brief information of
these events is presented in Appendix~\ref{AppendixB}.  The posterior samples
are obtained from the Gravitational Wave Open Science
Center (GWOSC)\footnote{\url{https://gwosc.org/}}~\cite{LIGOScientific:2019lzm,
KAGRA:2023pio, LIGOScientific:2025snk}.  We evaluate the reliability of our two encoded data
products---high-fidelity neural network weights and analytic expressions---by
comparing the posteriors of raw samples with those of the resampled samples.
Additionally, we further ensure the reliability of encoded data products in
practical applications by performing population inference using reconstructed
posterior samples.

For the chosen GW posteriors, a key challenge for us is that some parameter
distributions are non-smooth, being truncated at their prior boundaries; for
example, ${\cal M} > 0$ and $q \leq 1$.  Direct density estimation on such
truncated distributions would require substantially more network parameters to
capture these sharp edges.  To address this, we apply a data-driven
preprocessing step that reparameterizes each truncated distribution into an
untruncated form defined over the full parameter space, easing the estimation
task and potentially enhancing the model's robustness.  Specifically, we remove
the bounds on the parameters using invertible transformations.  Because the
chirp mass is limited to be positive, we apply a transform through
%--
\begin{equation}
    \ln{\mathcal{M}}' = \frac{\ln{\mathcal{M}} -
    \mu[\ln{\mathcal{M}}]}{\sigma[\ln{\mathcal{M}}]}\, ,
\label{eq:7}
\end{equation}
%--
where $\mu[\ln{\mathcal{M}}]$ and $\sigma[\ln{\mathcal{M}}]$ are the mean and
standard deviation of the  logarithm of the chirp mass respectively.  Among the
remaining parameters, the mass ratio and spin magnitudes are constrained within
the interval $[0,1]$, while the spin inclinations are restricted to $[0,\pi]$.
We follow~\citet{Hoy:2020vys} and apply the logit transformation to these
parameters,
%--
\begin{equation}
    \theta'=\ln\left(\frac{\theta-a}{b-\theta}\right)\, .
\label{eq:8}
\end{equation}
%--
Here, $\theta$ represents the original parameter value, and $a$ and $b$ are the
lower and upper bounds of the parameter, respectively.  All these
transformations are invertible and performed before training the neural network.
After the training phase, inverse transformations are performed during the
sampling phase to recover the original parameter values.  Other methods such as
boundary reflection can also be used to handle the bounded
parameters~\cite{jones1993simple}.

After data preprocessing, neural network training, and symbolification, we
obtain two encoded representations of the posterior samples for each event.
For these ten GW events, all neural networks adopt a compact two-layer
architecture. We empirically explored different choices for the number of Gaussian components
$G$ and B-spline basis functions $B$. Increasing $G$ and $B$ beyond moderate values
yielded only marginal gains, whereas reducing them degraded reconstruction fidelity.
Accordingly, we selected $G=5$ and $B=23$ as a practical compromise between accuracy and compactness.
Taking GW150914 as an example, Tab.~\ref{table:1} lists
the storage requirements and sampling rates for both types of encoded data
products. The neural network weights and analytic expressions achieve
approximately $90\times$ and $3500\times$ compression ratios in storage,
respectively.

\begin{figure}[]
    \includegraphics[width=\linewidth]{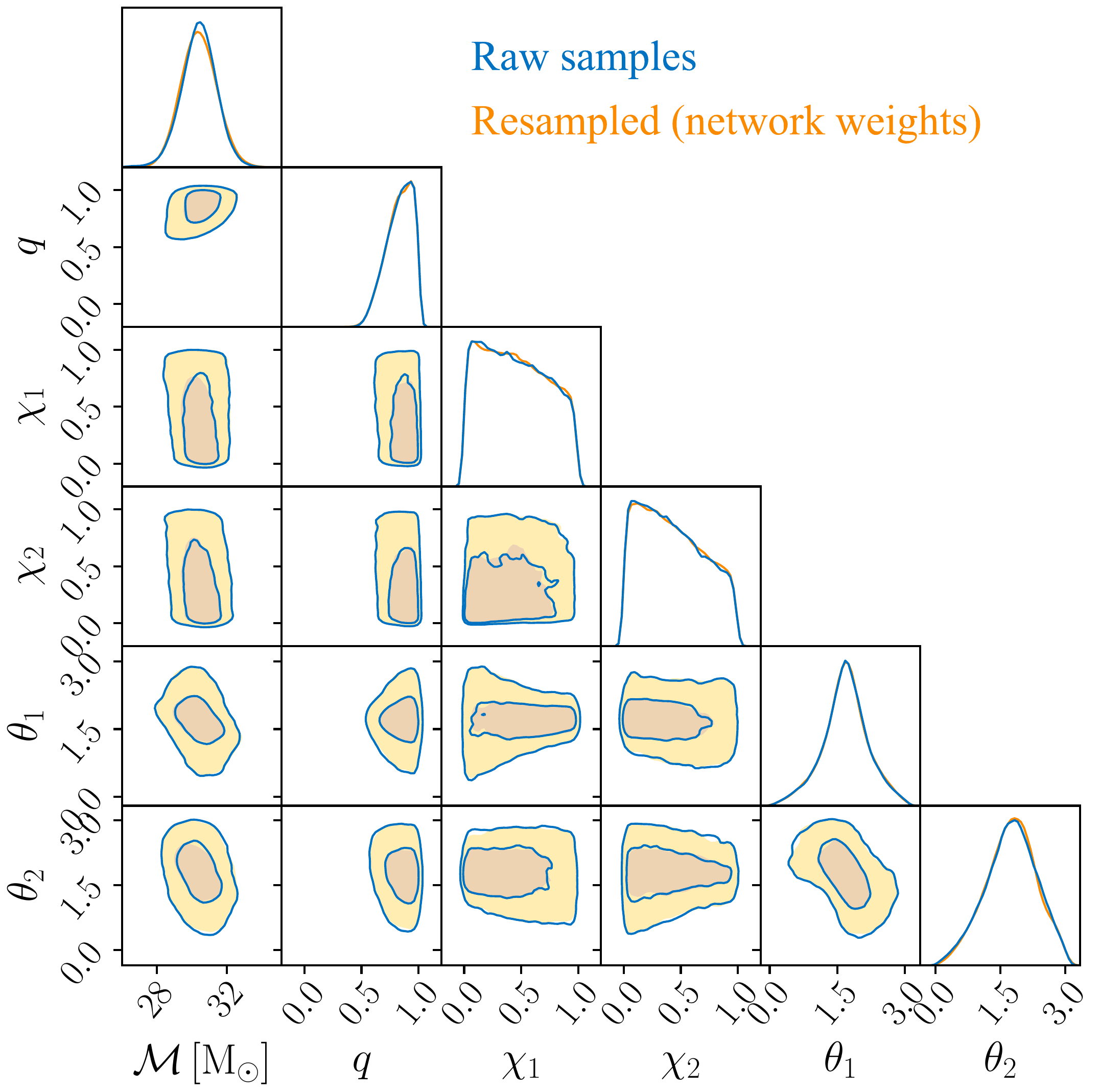}
    \caption{Marginalized one- and two-dimensional distributions, comparing
    samples generated with the neural network weights (orange) and raw samples
    (blue) of GW150914.  Both of them contain 147,634 samples.  Contour lines in
    the two-dimensional joint distributions delineate the $50\%$ and $90\%$
    credible regions.}
    \label{fig:5}
\end{figure}

\begin{figure}[]
    \includegraphics[width=\linewidth]{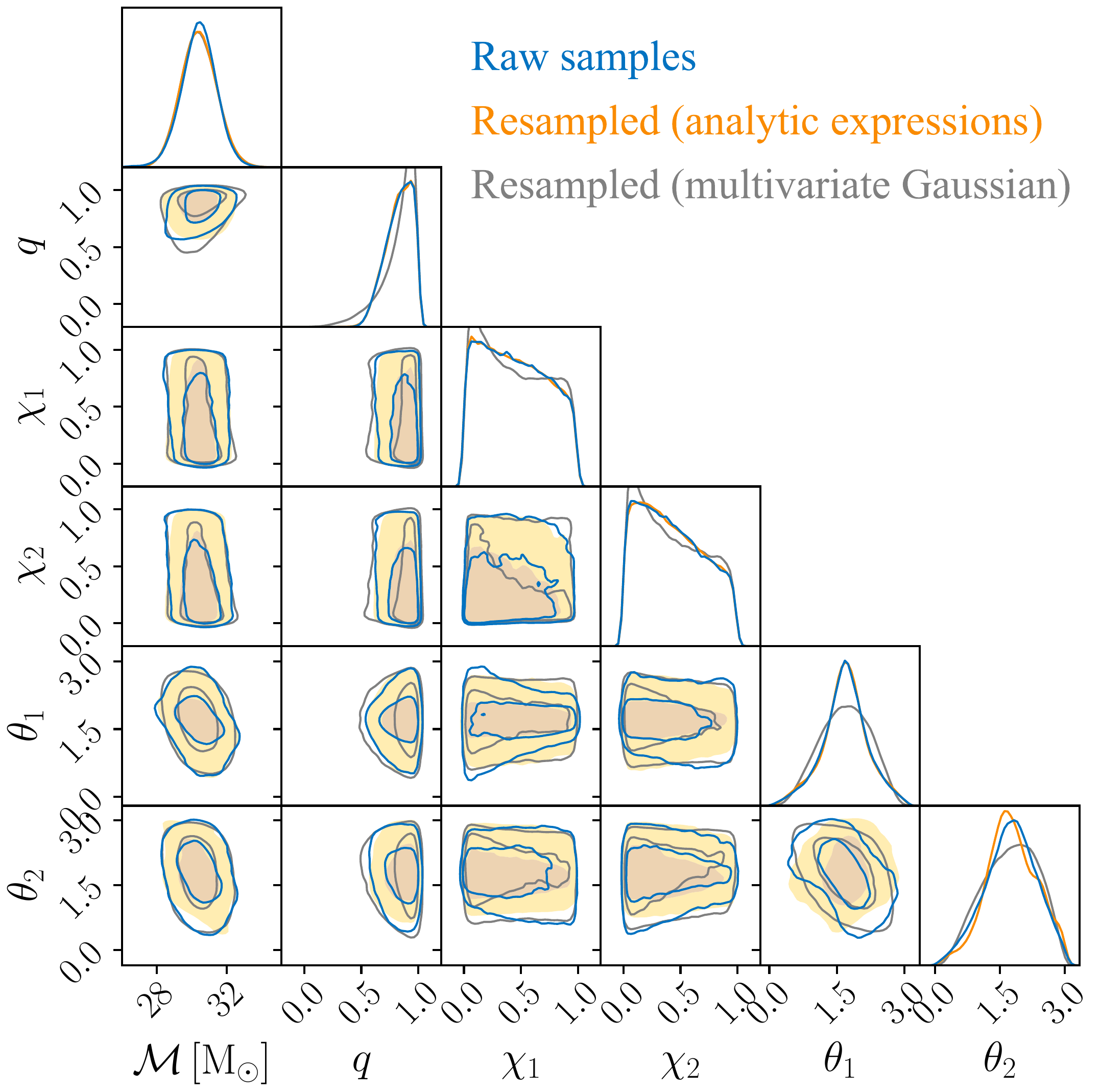}
    \caption{Same as Fig.~\ref{fig:5}, but for samples generated with analytic
    expressions (orange), the fitted multivariate Gaussian distribution (grey),
    and raw samples (blue).}
    \label{fig:6}
\end{figure}

The resampling results are shown in Fig.~\ref{fig:5} and Fig.~\ref{fig:6}.  As
shown in Fig.~\ref{fig:5}, the samples generated from the neural network weights
exhibit close agreement with the original distribution.
In Fig.~\ref{fig:6}, samples generated with the
analytic expressions capture the overall structure of the posterior
distribution, but exhibit some deviations from the original distributions.
Compared with the neural network weights, this reflects some trade-off between
fidelity and the gains in storage efficiency, interpretability, and expression
simplicity.  By replacing the network's nonlinear edge functions with more
restrictive analytical forms, the model's complexity is inevitably reduced and
some fidelity in capturing local features is lost.  The analytic expression data
products occupy only $\mathcal{O}\,(1\,{\rm KB})$ of storage.  In
Fig.~\ref{fig:6} we also show samples from a fitted multivariate Gaussian
distribution, where we use the same data preprocessing method as in the
KAN-based neural density estimator. In preparing them, we start from the
initial mean and covariance matrix of the preprocessed raw samples, and obtain
the optimal mean and covariance matrix of the 6-dimensional Gaussian
distribution by minimizing the average negative log-posterior.  Compared to
other representations, such as summarizing GW posteriors by their mean and
covariance matrix, as well as its higher-order extension~\cite{Wang:2022kia},
the analytic expressions produced by the method of this work are much better at
capturing the distortions and multimodal features of GW posterior distributions,
giving a better balance between accuracy and storage efficiency.  The
probability-probability (P-P) plot comparing the raw and resampled samples
of GW150914 is shown in Fig.~\ref{fig:4}. The resampling results for other nine
GW events are shown in Appendix.~\ref{AppendixB}.

\begin{figure}[]
    \includegraphics[width=\linewidth]{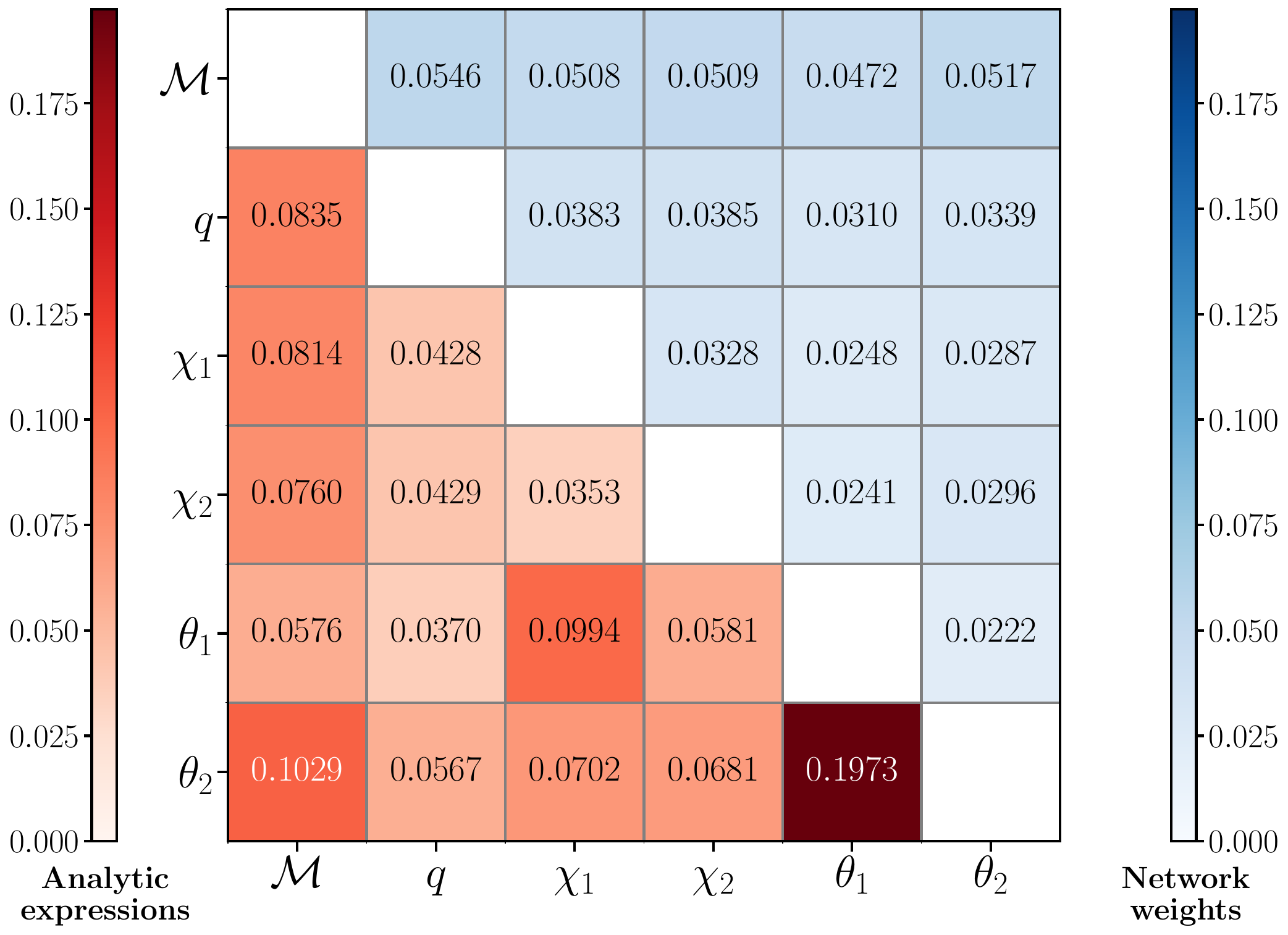}
    \caption{{\bf Two-dimensional Hellinger distances for pairwise combinations of the
    parameters of GW150914.} The distances take values in $[0,1]$, with smaller values
    indicating closer agreement. The top right corner shows the comparison between original
    samples and samples generated using network weights. The bottom left corner shows the
    comparison between original samples and samples generated using analytic expressions.}
    \label{fig:7}
\end{figure}

To quantify the differences between the distributions of the raw and resampled GW150914
posterior samples, we calculate the Wasserstein distances~\cite{1969Markov,computational2019Peyre}
of the six one-dimensional marginalized distributions between raw and reproduced posteriors.
Wasserstein distance is a distance function defined between two
probability distributions that measure the similarity of them.
It is worth noting that this distance metric characterizes differences 
in one-dimensional marginal distributions, 
while corner plots provide a complementary way to compare correlations among parameters. 
Compared with the more commonly used Kullback-Leibler divergence~\cite{Kullback:1951zyt},
Wasserstein distance is more universal and can reflect the similarity between two distributions which
overlap very little. Intuitively, if one regards each probability distribution
as an equal (normalized) pile of soil, then the Wasserstein distance can be
interpreted as the required minimum ``work'' to deform one pile of soil into another,
which is equal to the amount of soil to be moved multiplied by the average ``distance''
between two piles (distributions)~\cite{Wang:2022kia}. We normalize the Wasserstein
distances using the standard deviation of each one-dimensional marginalized distribution
in raw posteriors and the results are shown in Tab.~\ref{table:2}. Similar distributions will
have a small Wasserstein distance. We can see that most of the one-dimensional
marginalized distributions between raw samples and samples resampled with our data products
match very well with Wasserstein distances smaller than $0.02$, indicating that the difference
is less than $0.02$ standard deviation. Although samples generated with analytic expressions
show relatively larger deviation in the marginalized distribution of $\theta_2$, analytic expressions
outperform the multivariate Gaussian model in all six parameters. As a metric for
quantifying differences between high-dimensional distributions, we compute the two-dimensional
Hellinger distances~\cite{hellinger1909neue} for pairwise combinations of the parameters of GW150914.
The results, summarized in Fig.~\ref{fig:7}, provide a complementary assessment of the joint correlations.

Additionally, we apply the above KAN-based density estimator 
to the reconstruction and compression of the full 15-dimensional posterior for 
GW150914. The  data product based on network weights achieves a $34\times$ compression, 
demonstrating that the method remains robust and effective in the full-parameter 
scenario. Details of the experimental configuration and the reconstructed 
posteriors are presented in Appendix~\ref{AppendixC}.

%---------------------------------------------------------------------
\begin{table}[]
    \renewcommand\arraystretch{1.5}
    \caption{{\bf Comparison between two data-compression products for GW150914's
    posterior distribution.} ``$G$'' denotes the number of Gaussian components
    in GMMs that fit the conditional probability distribution~(\ref{eq:2});
    ``Hidden'' indicates the number of hidden layers in the neural network
    architecture before symbolification; ``Storage'' shows the storage space
    ratio of data products to raw samples, with the raw data containing 147,634
    six-dimensional posterior samples in HDF5 format; ``Sampling Time''
    represents the average re-sampling duration of 20 runs with 2,000 samples
    per run. Averaged over five runs, network training took 27.14 seconds
    on the same hardware as in resampling.}
    \resizebox{\linewidth}{!}{
    \begin{tabular}{lcccc}
    \hline\hline
    Type & $G$ & Hidden & Storage (KB) & Sampling Time\footnote{NVIDIA RTX 4070 Laptop GPU (8 GB)}  (s) \\
    \hline
    Network Weight & 5 & 0 & 77/6921 & 0.8809 \\
    Analytic Expression & 5 & 0 & 2.1/6921 & 0.0900 \\
    \hline
    \end{tabular}
    }
    \label{table:1}
\end{table}
%---------------------------------------------------------------------
\begin{table}[]
    \renewcommand\arraystretch{1.5}
        \caption{{\bf Normalized one-dimensional Wasserstein distances for the
    posteriors of three data products.} Three rows from top to the bottom show the
    comparisons between raw samples and the samples generated using neural
    network weights, analytic expressions and fitted multivariate Gaussian
    distribution, respectively. Each element in the table represents the
    normalized Wasserstein distance between the one-dimensional marginal
    distributions.}
    \resizebox{\linewidth}{!}{
    \begin{tabular}{lcccccc}
    \hline\hline
    Type & $\mathcal{M}$ & $q$ & $\chi_1$ & $\chi_2$ & $\theta_1$ & $\theta_2$ \\
    \hline
    Network Weight & 0.051 & 0.009 & 0.008 & 0.017 & 0.006 & 0.019 \\ 
    Analytic Expression & 0.050 & 0.011 & 0.005 & 0.006 & 0.010 & 0.062 \\ 
    Multivariate Gaussian & 0.052 & 0.261 & 0.044 & 0.045 & 0.134 & 0.075 \\
    \hline
    \end{tabular}
    }
  \label{table:2}
\end{table}
%---------------------------------------------------------------------
%---------------------------------------------------------------------
%---------------------------------------------------------------------
\begin{figure*}[]
    \includegraphics[width=0.8\linewidth]{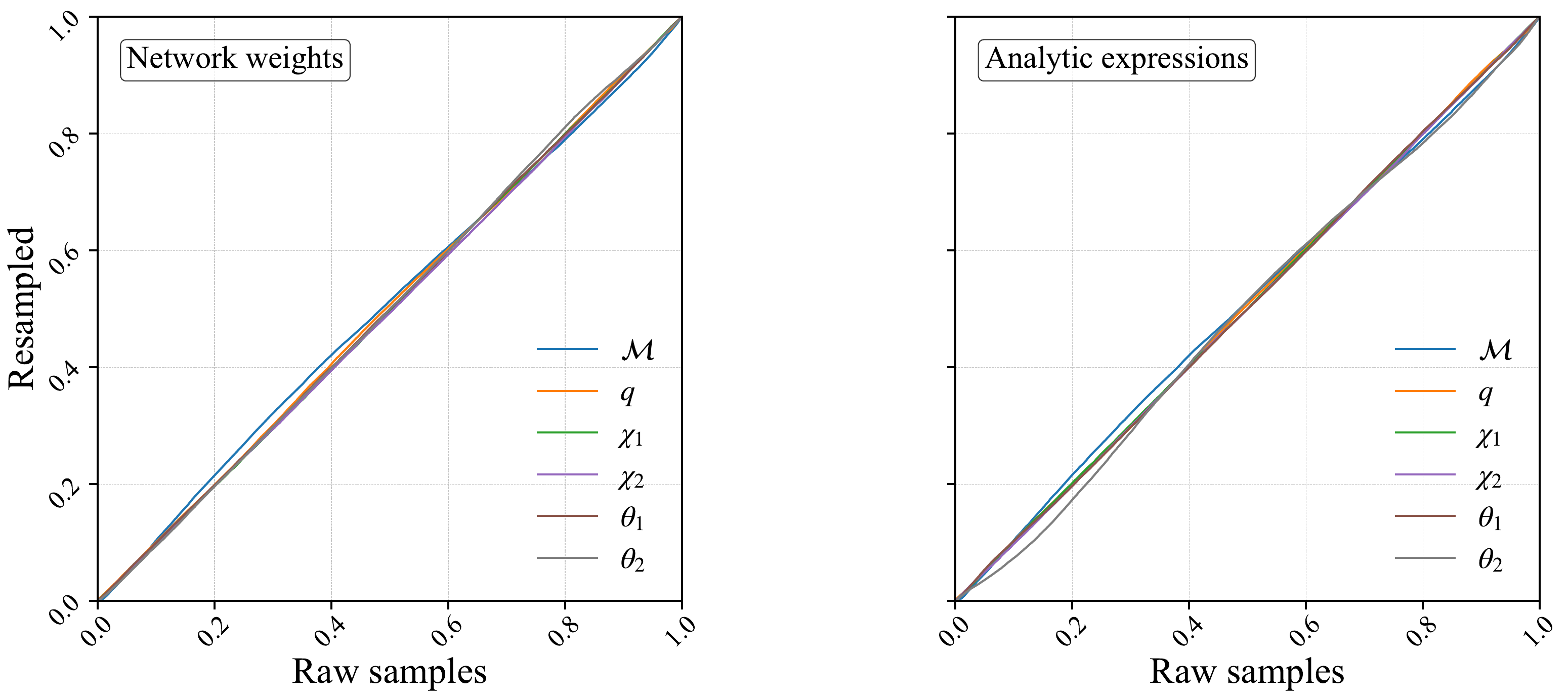}
    \caption{P-P plot comparing the raw posterior samples of GW150914 with those
    resampled with the trained neural network (left) and fitted analytic
    expressions (right). }
    \label{fig:4}
\end{figure*}
%---------------------------------------------------------------------
%---------------------------------------------------------------------

Many key scientific results in GW astronomy, such as constraints on the mass and
spin distributions of black holes~\cite{Dong:2024zzl,Tong:2022iws}, rely on
population inference that combines posterior samples from many events.  In the
era of next-generation GW detectors, the number of detected events is expected
to increase explosively, and the demand for posterior samples in population
studies will grow accordingly.  This will pose challenges on users for both
data storage and transmission.  The primary motivation for proposing a
lightweight posterior catalog is to address the need for efficient storage and
distribution of large-scale GW posterior data in the future.  To ensure that our
method preserves not only event-level fidelity but also the statistical
properties required for population-level scientific applications, we further
validate the reliability of our lightweight data products by performing
population inference using samples resampled using our two types of compact data products.

In population inference, we seek to determine the posterior distribution for the
population parameters (or hyperparameters), $\boldsymbol{H}$, given the
collective data $\mathcal{D}=\big\{d_i \big\}_{i=1}^N$ from $N$ independent
events.  According to Bayes' theorem, this posterior, denoted
$P(\boldsymbol{H}|\mathcal{D})$, is expressed as, 
%---
\begin{equation}
    P(\boldsymbol{H}|\mathcal{D}) = \frac{P(\mathcal{D}|\boldsymbol{H})
    P(\boldsymbol{H})}{P(\mathcal{D})}\, ,
    \label{eq:9}
\end{equation}
%---
where $P(\mathcal{D}|\boldsymbol{H})$ is the total marginalized likelihood over
the source parameters of the events and $P(\boldsymbol{H})$ is the prior
distribution of the population parameters.  Using posterior samples in the GW
catalog, $P(\mathcal{D}|\boldsymbol{H})$ is computed as
follows~\cite{thrane2019introduction},
%---
\begin{equation}
    P(\mathcal{D}|\boldsymbol{H}) = \prod_{i=1}^{N}\frac{\mathcal{Z}(d_i  |
    \emptyset)}{n_i} \sum_{k=1}^{n_i}
    \frac{\pi(\boldsymbol{\theta}_i^k|\boldsymbol{H})}{\pi(\boldsymbol{\theta}_i^k|\emptyset)}\,
    .
    \label{eq:10}
\end{equation}
%---
Here, the summation is over the $n_i$ posterior samples,
$\{\boldsymbol{\theta}_i^k\}$, for event $i$.  The
term $\pi(\boldsymbol{\theta}_i^k|\emptyset)$ denotes some default prior
in the original analysis and $\mathcal{Z}$ denotes the corresponding evidence,
while $\pi(\boldsymbol{\theta}_i^k|\boldsymbol{H})$ represents
prior in the population model, controlled by the hyperparameters $\boldsymbol{H}$.

\begin{figure*}[]
    \centering
    \includegraphics[width=0.8\linewidth]{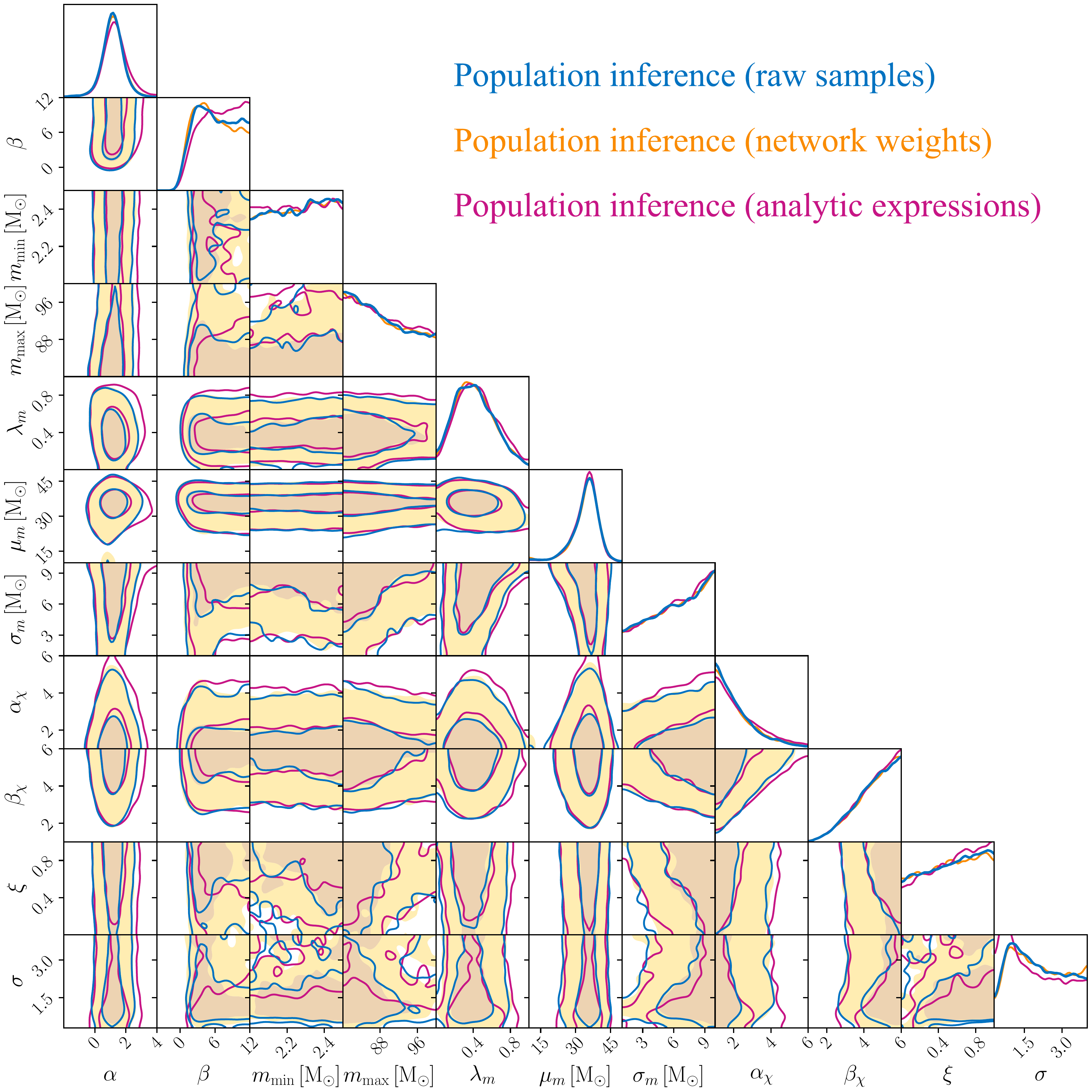}
    \caption{Marginalized one- and two-dimensional distributions, comparing the
    population inference results using raw samples (blue), samples generated
    with our neural networks (orange) and analytic expressions (pink) of 10 GW events.
    Contour lines in the two-dimensional joint distributions delineate the $50\%$ and $90\%$ credible
    regions.  The sampling is accomplished using the {\it dynesty sampler}~\cite{Speagle:2019ivv}
    and {\it nlive} is set to 2,000.}
    \label{fig:8}
\end{figure*}

We compare the posterior distributions, $P(\boldsymbol{H}|\mathcal{D})$, based
on three sets of samples: the raw samples for each event from GWOSC, the
recovered samples using corresponding neural network weights and the recovered
samples using corresponding analytic expressions. All three sets contain the
same 10 events, and for each event the number of posterior samples equals to that
of the raw samples. We use the \texttt{GWPopulation} package~\cite{Talbot:2024yqw}
to perform population inference on these three sets of samples. The analyses use the
{\it two-component-primary-mass-ratio} and {\it iid-spin} models, with detailed
description and prior choices of the hyperparameters given in
Appendix~\ref{AppendixD}.  Comparative results for the three population-level
posterior distributions are shown in Tab.~\ref{table:3} and Fig.~\ref{fig:8}.
As shown in the figure, population inference results of raw samples
and samples recovered using neural network weights match very well, validating
the feasibility of our compression strategy. Result of samples recovered using
analytic expressions exhibits slight deviations. This behavior may in part stem
from the numerical instability of hierarchical Bayesian inference~\cite{Leyde:2023iof}, 
which can amplify small variations in the single-event posteriors, particularly for
weakly constrained hyperparameters such as the mass-ratio slope.
%---------------------------------------------------------------------
%---------------------------------------------------------------------
\begin{table*}[]
    \renewcommand\arraystretch{1.5}
    \caption{{\bf Normalized one-dimensional Wasserstein distances for the population posteriors.}
    The two rows show the comparisons between the population inference results
    using raw samples and the samples generated using neural network weights and
    the samples generated using analytic expressions, respectively.}
    \begin{tabularx}{\linewidth}{p{2.5cm}p{1.3cm}p{1.3cm}p{1.3cm}p{1.3cm}p{1.3cm}p{1.3cm}p{1.3cm}p{1.3cm}p{1.3cm}p{1.3cm}p{1.3cm}}
    \hline\hline
    Type & $\alpha$ & $\beta$ & $m_{\rm min}$ & $m_{\rm max}$ & $\lambda_m$ & $\mu_m$ & $\sigma_m$ & $\alpha_{\chi}$ & $\beta_{\chi}$ & $\xi$ & $\sigma$\\
    \hline
    Network Weight & 0.018 & 0.032 & 0.027 & 0.028 & 0.049 & 0.012 & 0.022 & 0.030 & 0.015 & 0.014 & 0.018\\
    Analytic Expression & 0.199 & 0.225 & 0.032 & 0.069 & 0.169 & 0.028 & 0.046 & 0.165 & 0.012 & 0.064 & 0.215\\
    \hline
    \end{tabularx}
    \label{table:3}
\end{table*}
%---------------------------------------------------------------------

%---------------------------------------------------------------------
\section{Summary}
\label{Sec5_Summary}
%---------------------------------------------------------------------

In this work, we propose a KAN-based neural density estimator and explore its
potential for constructing high-fidelity, low-storage surrogate GW posterior
catalogs.  By harnessing the flexibility of KAN's learnable spline activations,
our estimator delivers enhanced interpretability and parameter efficiency.  We
first demonstrate its analytic-fitting capabilities on verifiable examples and
then apply it to real GW posteriors.  With a KAN-based estimator, the posterior
samples are compactly represented as high-fidelity neural network weights with
$\mathcal{O}\,(10^2)$-fold compression, and analytic expressions with
$\mathcal{O}\,(10^3)$-fold compression, maintaining a balance between fidelity
and storage footprint. Once these data products are prepared, users can download
these encoded files to flexibly and rapidly resample posterior samples, greatly
reducing data-storage and data-transfer overhead.  We also validate the fidelity
of the compressed data products in a hierarchical population inference,
confirming their reliability for downstream high-level analyses.

Our work provides a preliminary demonstration of the parameter efficiency of KAN
in density estimation tasks.  To the best of our knowledge, this is the first
neural density estimator capable of yielding analytic expressions for the PDF.
While certain challenges remain in modeling highly complex distributions, the
analytic PDF representations obtained by our method have shown clear advantages
in storage efficiency and sampling speed.  In future work, we plan to further
evaluate the performance of the KAN-based neural density estimator in broader
application scenarios, such as in GW posterior estimation with conditional
networks.  This strategy for lightweight GW posterior catalog construction could
bring inspiration for tackling the data challenges and pressures of
next-generation GW detectors, and it shows how advances in neural network
architectures can further empower GW data science.

%---------------------------------------------------------------------
\section*{Acknowledgements}

We thank the anonymous referee for comments.
This work was supported by the Beijing Natural Science Foundation (1242018), the
National Natural Science Foundation of China (123B2043, 12573042), the National SKA
Program of China (2020SKA0120300), the Max Planck Partner Group Program funded
by the Max Planck Society, and the High-performance Computing Platform of Peking
University. 

This research has made use of data or software obtained from the Gravitational
Wave Open Science Center (gwosc.org), a service of the LIGO Scientific
Collaboration, the Virgo Collaboration, and KAGRA. This material is based upon
work supported by NSF's LIGO Laboratory which is a major facility fully funded
by the National Science Foundation, as well as the Science and Technology
Facilities Council (STFC) of the United Kingdom, the Max-Planck-Society (MPS),
and the State of Niedersachsen/Germany for support of the construction of
Advanced LIGO and construction and operation of the GEO600 detector. Additional
support for Advanced LIGO was provided by the Australian Research Council. Virgo
is funded, through the European Gravitational Observatory (EGO), by the French
Centre National de Recherche Scientifique (CNRS), the Italian Istituto Nazionale
di Fisica Nucleare (INFN) and the Dutch Nikhef, with contributions by
institutions from Belgium, Germany, Greece, Hungary, Ireland, Japan, Monaco,
Poland, Portugal, Spain. KAGRA is supported by Ministry of Education, Culture,
Sports, Science and Technology (MEXT), Japan Society for the Promotion of
Science (JSPS) in Japan; National Research Foundation (NRF) and Ministry of
Science and ICT (MSIT) in Korea; Academia Sinica (AS) and National Science and
Technology Council (NSTC) in Taiwan of China.

%---------------------------------------------------------------------
\textbf{Code Availability.} The code used in this work is publicly available at
\url{https://github.com/liu-ws/KMADE}.
%---------------------------------------------------------------------

\appendix
\section{A Case Test Beyond Gaussian Families}

\label{AppendixA}

\begin{figure}[]
    \centering
    \includegraphics[width=0.7\linewidth]{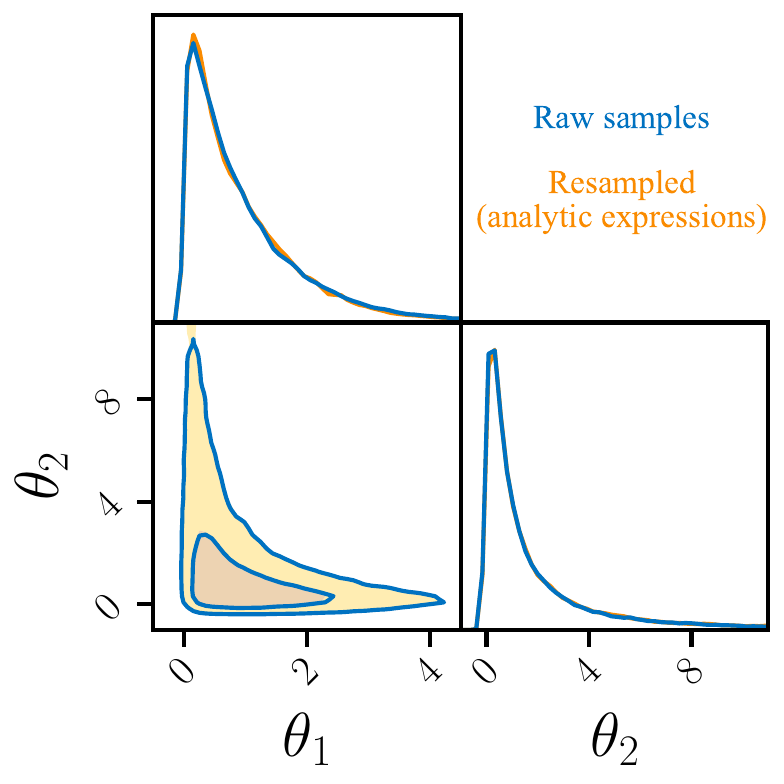}
    \caption{{\bf Results of additional illustrative example.} Marginalized one- and
    two-dimensional distributions are shown in corner plots, comparing samples
    resampled from fitted expressions (orange) and samples generated from the
    ground-truth distribution (blue).  Both of them contain 40,000 samples.}
    \label{fig:9}
\end{figure}

In this appendix, we present an additional illustrative example beyond Gaussian
families to more comprehensively demonstrate the effectiveness of our method
in modeling diverse distributions. In Sec.~\ref{Sec3_Experiments}, we chose 
three cases parameterized by Gaussian or Gaussian mixture distributions because
they provide closed-form ground truths. It is worth noting that our method 
remains effective beyond Gaussian families. To illustrate this,
we present an additional example using an exponential-type distribution:
%--
$$
P(\theta_1,\theta_2)=\frac{1}{\theta_1}e^{-\theta_1}e^{-\frac{\theta_2}{\theta_1}}.
$$
%--
We fit this target with three mixture components per conditional and obtain the following
learned expression,
$$
\begin{aligned}
P'(\theta_1,\theta_2)=&\frac{1}{2\pi}[ 0.58e^{-0.63(\theta_1+0.75)^2}+0.60e^{-1.45(\theta_1-0.37)^2}\\
&+0.09e^{-1.32(0.41\theta_1+1.00)^2} ]\times[0.86e^{-0.78(0.97\theta_1+\theta_2+0.03)^2}\\
&+0.24e^{-0.77(0.65\theta_1+0.63\theta_2+1.00)^2}+0.00].
\end{aligned}
$$
The reconstructed
distribution is visualized in Fig.~\ref{fig:9}.
This example demonstrates that, in practice, our estimator can recover the 
target distribution with high fidelity, 
serving as an additional validation of the method's effectiveness.

%---------------------------------------------------------------------
\section{GW Events and Sampling Results}
\label{AppendixB}
%---------------------------------------------------------------------

In this appendix, we give some information of binary black hole GW events in
GWTC-1 (GW150914, GW151012, GW151226, GW170104, GW170608, GW170729, GW170809,
GW170814, GW170818, GW170823~\cite{LIGOScientific:2018mvr}), and the resampling
results obtained with our compact data products.

All ten GW events used in this paper are binary black hole mergers composed of
stellar-mass black holes, originally detected during the LIGO/Virgo's first and
second observing runs~\cite{LIGOScientific:2018mvr}.  The raw posterior samples
used in this paper are taken from the GWTC-2.1
catalog~\cite{LIGOScientific:2021usb}, which features a reanalysis of these
events with state-of-the-art waveform models.  This population covers a range of
total masses from approximately $18.5\,M_{\odot}$ (GW170608) to
$84.4\,M_{\odot}$ (GW170729).  While most systems have comparable component
masses, the spin characteristics show important variations.  The majority of
events have effective inspiral spins ($\chi_{\rm eff}$) consistent with zero.
However, GW151226 stands out with a measured positive spin, providing evidence
for at least one spinning black hole in that system.  Furthermore, parameter
uncertainties vary significantly across the events, often correlating with
detection significance: the inaugural loud event GW150914 with a network SNR of
$26$ has well-constrained properties, whereas the most distant event GW170729 at
redshift $z \approx 0.44$ exhibits much broader posterior distributions due to
its lower SNR.

\begin{figure*}[]
    \centering
    \includegraphics[width=\linewidth]{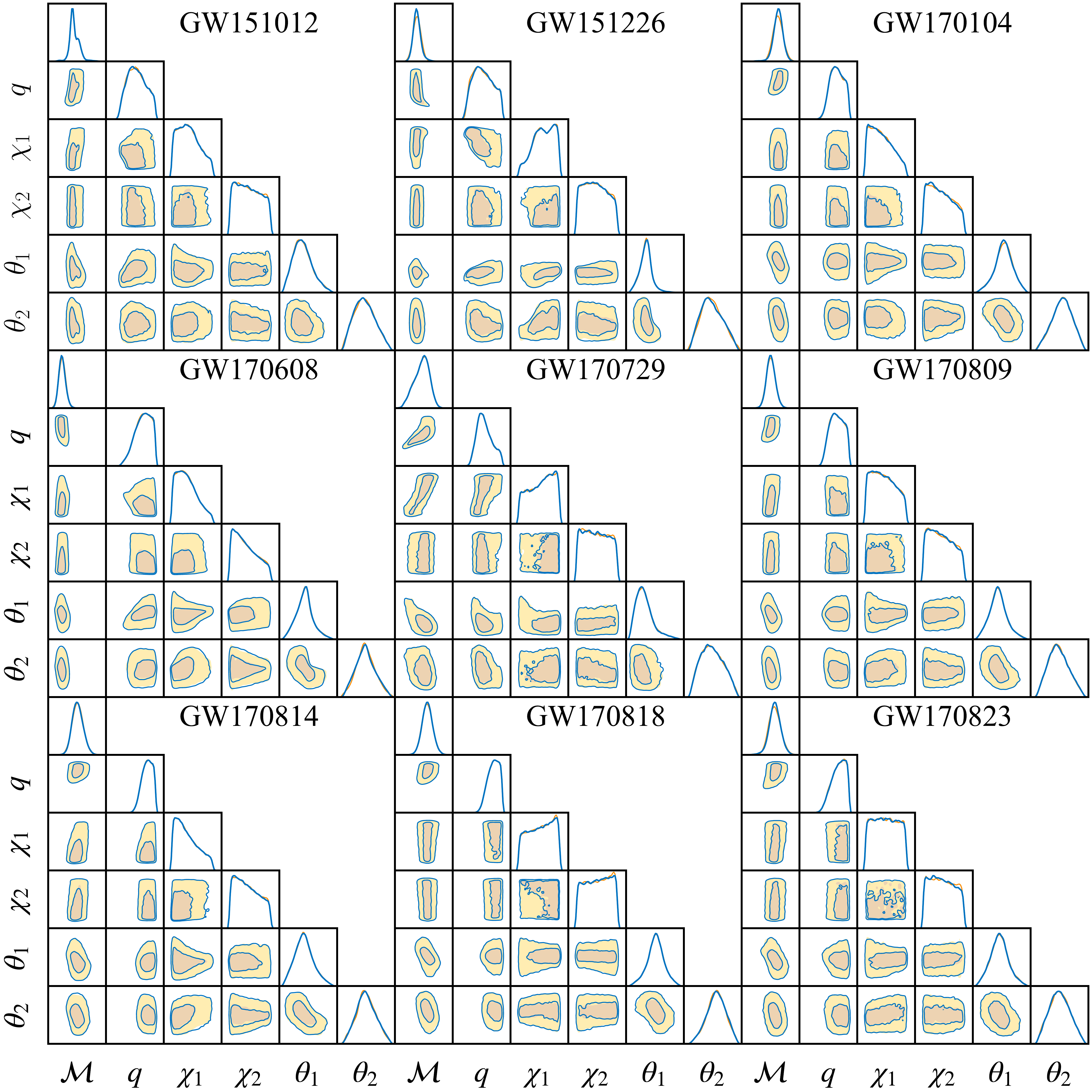}
    \caption{Marginalized one- and two-dimensional distributions, comparing
    samples generated with the neural network weights (orange) and raw samples
    (blue) of other nine events.  Contour lines in the two-dimensional joint
    distributions delineate the $50\%$ and $90\%$ credible regions.}
    \label{fig:10}
\end{figure*}

\begin{figure*}[]
    \centering
    \includegraphics[width=\linewidth]{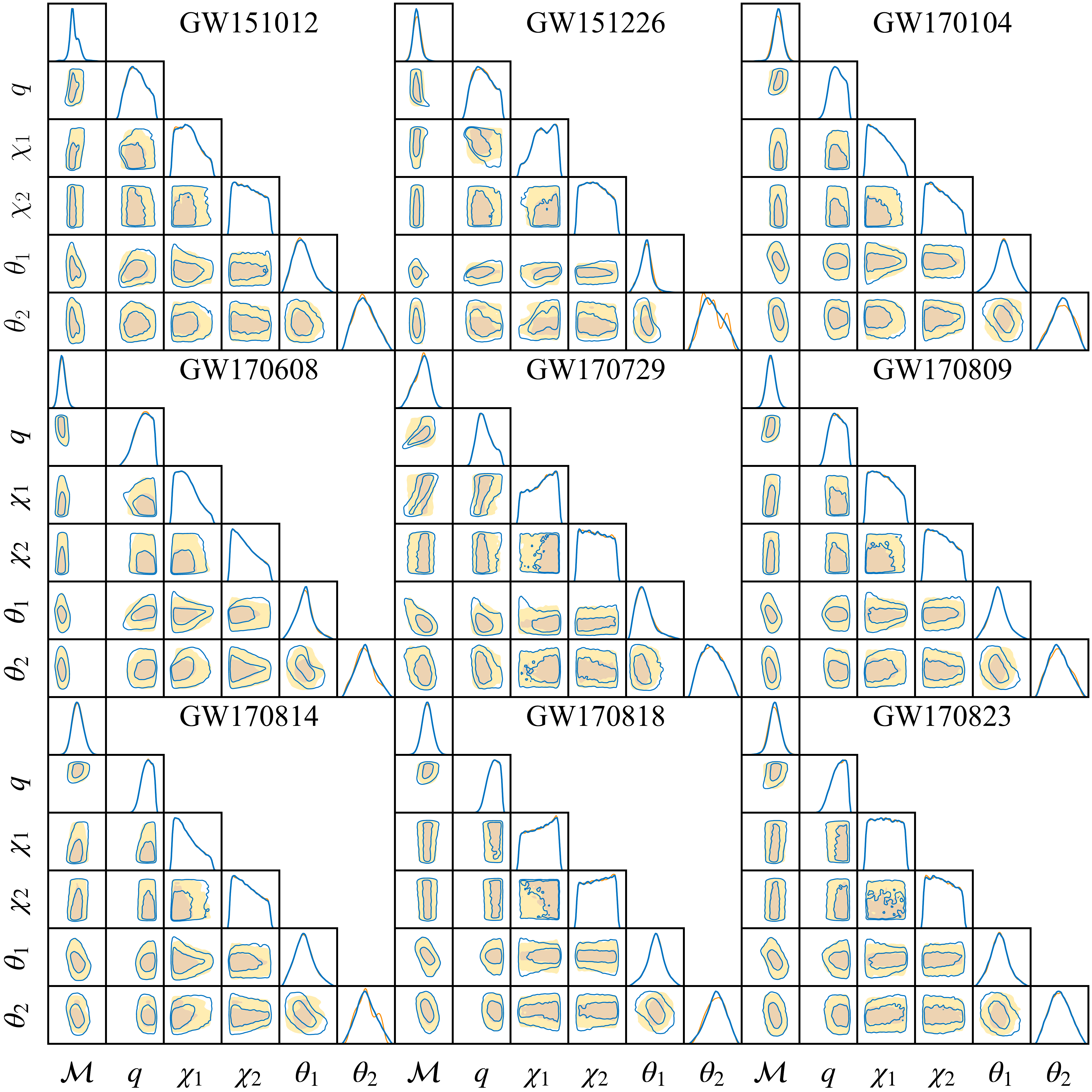}
    \caption{Same as Fig.~\ref{fig:10}, but for samples generated with analytic
    expressions (orange) and raw samples (blue).}
    \label{fig:11}
\end{figure*}

The sampling results are shown in Fig.~\ref{fig:10} and Fig.\ref{fig:11} and
the normalized one-dimensional Wasserstein distances are listed in Tab.~\ref{table:4}.
As shown in Fig.~\ref{fig:10}, the samples generated from the neural network weights
exhibit close agreement with the original distribution from GWOSC.
In Fig.~\ref{fig:11}, samples generated with the
analytic expressions capture the overall structure of the posterior
distribution, but exhibit some deviations from the original distributions.
In producing the results in the figures, the
data preprocessing method, neural network configurations, training processes,
and training devices of all ten events are kept consistent.  During resampling,
the number of samples for each event is the same as that of the raw samples, and
the resampling device also remains consistent.

%---------------------------------------------------------------------
%---------------------------------------------------------------------
%---------------------------------------------------------------------
\begin{table*}[]
    \renewcommand\arraystretch{1.5}
        \caption{{\bf Normalized one-dimensional Wasserstein distances for the
    posteriors of data products of other nine events.} Similar to Tab.~\ref{table:2}.}
    \begin{tabularx}{\linewidth}{p{2cm}p{4cm}p{2cm}p{2cm}p{2cm}p{2cm}p{2cm}p{2cm}}
    \hline\hline
    Event & Type & $\mathcal{M}$ & $q$ & $\chi_1$ & $\chi_2$ & $\theta_1$ & $\theta_2$ \\
    \hline
    GW151012 & Network Weight & 0.012 & 0.007 & 0.006 & 0.008 & 0.016 & 0.020 \\ 
    & Analytic Expression & 0.007 & 0.011 & 0.025 & 0.005 & 0.016 & 0.019 \\
    \hline
    GW151226 & Network Weight & 0.118 & 0.025 & 0.010 & 0.019 & 0.016 & 0.036 \\
    & Analytic Expression & 0.116 & 0.010 & 0.004 & 0.004 & 0.074 & 0.053 \\
    \hline
    GW170104 & Network Weight & 0.121 & 0.014 & 0.014 & 0.036 & 0.021 & 0.011 \\
    & Analytic Expression & 0.120 & 0.003 & 0.005 & 0.006 & 0.008 & 0.069 \\
    \hline
    GW170608 & Network Weight & 0.066 & 0.015 & 0.008 & 0.014 & 0.014 & 0.044 \\
    & Analytic Expression & 0.067 & 0.018 & 0.003 & 0.004 & 0.020 & 0.094 \\
    \hline
    GW170729 & Network Weight & 0.067 & 0.007 & 0.007 & 0.011 & 0.012 & 0.009 \\
    & Analytic Expression & 0.022 & 0.016 & 0.006 & 0.007 & 0.036 & 0.023 \\
    \hline
    GW170809 & Network Weight & 0.034 & 0.018 & 0.008 & 0.013 & 0.012 & 0.015 \\
    & Analytic Expression & 0.033 & 0.007 & 0.008 & 0.003 & 0.012 & 0.051 \\
    \hline
    GW170814 & Network Weight & 0.031 & 0.007 & 0.004 & 0.015 & 0.010 & 0.026 \\
    & Analytic Expression & 0.030 & 0.005 & 0.007 & 0.010 & 0.008 & 0.047 \\
    \hline
    GW170818 & Network Weight & 0.024 & 0.007 & 0.009 & 0.009 & 0.012 & 0.024 \\
    & Analytic Expression & 0.022 & 0.011 & 0.005 & 0.006 & 0.008 & 0.046 \\
    \hline
    GW170823 & Network Weight & 0.074 & 0.012 & 0.004 & 0.024 & 0.023 & 0.012 \\
    & Analytic Expression & 0.076 & 0.009 & 0.005 & 0.010 & 0.014 & 0.010 \\
    \hline
    \end{tabularx}
  \label{table:4}
\end{table*}
%---------------------------------------------------------------------
\section{A Full 15-dimensional Posterior Reconstruction for GW150914}
\label{AppendixC}
%---------------------------------------------------------------------

\begin{figure*}[]
    \centering
    \includegraphics[width=\linewidth]{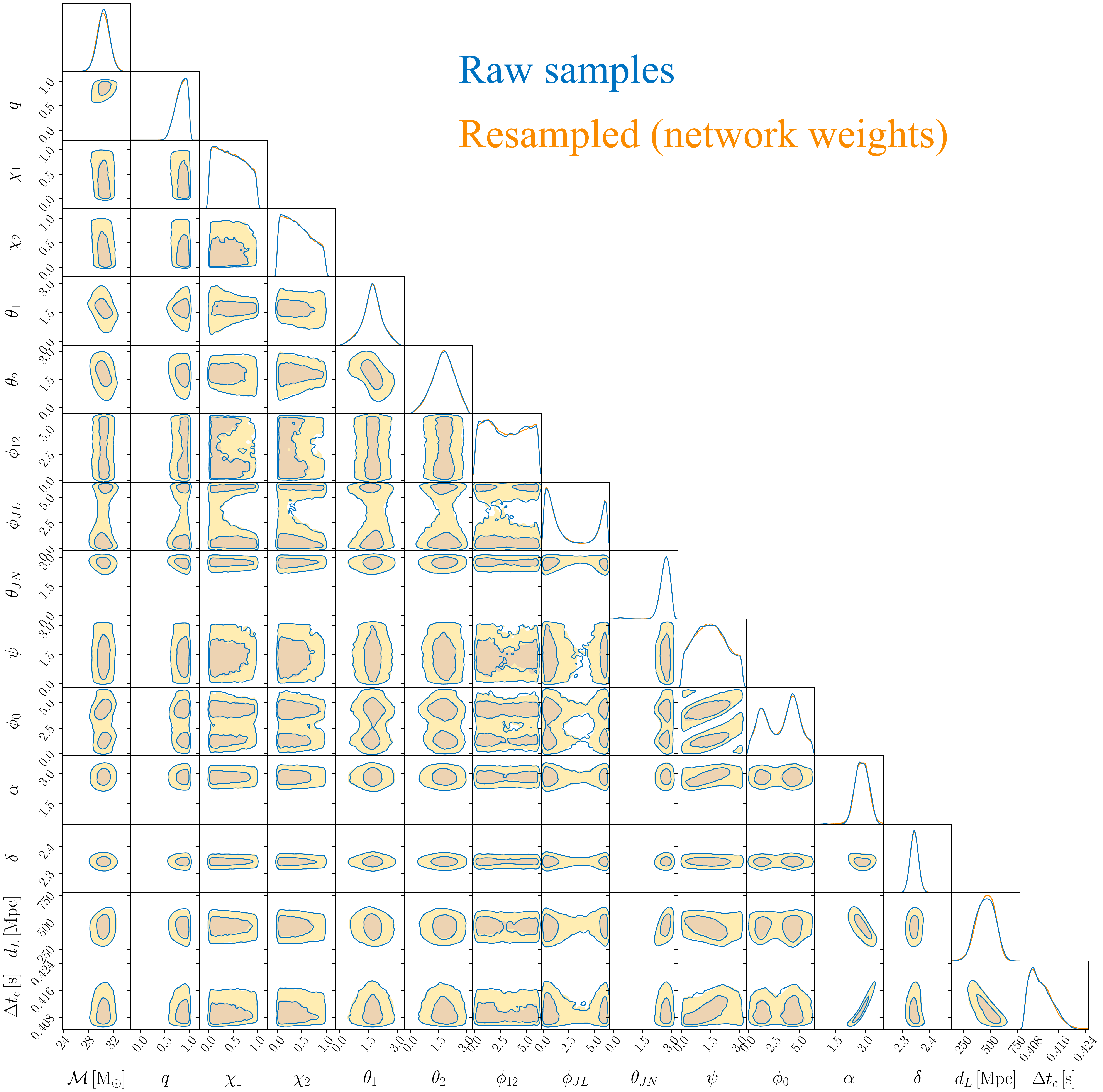}
    \caption{Marginalized one- and two-dimensional distributions, comparing
    samples generated with the neural network weights (orange) and raw samples
    (blue) for the full 15-dimensional posterior of GW150914. Contour lines in the two-dimensional 
    joint distributions delineate the $50\%$ and $90\%$ credible regions. In this figure, we have used $\Delta t_c \equiv t_c - 1126259462$\,s.}
    \label{fig:12}
\end{figure*}

In this appendix, we present the reconstruction and compression results for the full
15-dimensional posterior distribution of GW150914 using the KAN-based density estimator. 
Apart from the input dimension, the network architecture is essentially the same as 
in the 6-dimensional case, namely a compact two-layer network. Each conditional density 
is modeled with 5 Gaussian components, and each unmasked edge contains 23 cubic B-spline basis functions.
Despite the increased dimensionality, the trained network remains compact. The stored network weights occupy only
about 0.5 megabytes, compared with the 16.9 megabytes required to store the original posterior samples,
corresponding to a $34\times$ compression ratio. The resulting reconstructed posterior distribution
is shown in Fig.~\ref{fig:12}, which exhibits close agreement with the original distribution,
demonstrating the effectiveness and scalability of our KAN-based density estimator.

%---------------------------------------------------------------------
\section{Population Model and Hyperparameters}
\label{AppendixD}
%---------------------------------------------------------------------

In this appendix, we provide detailed description and priors of the
hyperparameters used in our population inference.  We employ the {\it
two-component-primary-mass-ratio} and {\it iid-spin} models from the
\texttt{GWPopulation} package~\cite{Talbot:2024yqw} for population inference.
The {\it two-component-primary-mass-ratio} is a power law model for
two-dimensional mass distribution. It models the primary mass and the
conditional mass ratio as,
%--
\begin{equation*}
    \begin{aligned}
        P(m_1,q) = & P(m_1)P(q|m_1)\\
        &\propto\left [(1-\lambda_m)m_1^{-\alpha} + \lambda_m
        \exp{\left(-\frac{(m_1-\mu_m)^2}{2\sigma_m^2}\right)}\right ]\times
        q^{\beta}\, .
    \end{aligned}
\end{equation*}
%--
Here $m_1$ is the primary mass, and $q$ is the mass ratio.  The {\it iid-spin}
model assumes independently and identically distributed spins.  The magnitudes
are assumed to follow a Beta distribution and the orientations are assumed to
follow an isotropic and truncated half Gaussian mixture model,
%--
\begin{equation*}
    \begin{aligned}
    P(\chi_1,\chi_2,\theta_1,\theta_1)=& P(\theta_1,\theta_2)P(\chi_1,\chi_2)\\
    &\propto\left[\frac{(1-\xi)^2}{4} + \xi\prod_{i\in{1,2}}\mathcal{N}_{[-1,1]}
    (\theta_i;\mu=1,\sigma=\sigma) \right]\\   
    &\times\prod_{j\in[1,2]}{\rm Beta}(\chi_j;\alpha_{\chi},\beta_{\chi})\, .
    \end{aligned}
\end{equation*}
%--
Here $\chi_1$ and $\chi_2$ are the spin magnitudes of the two black holes, and $\theta_1$
and $\theta_2$ are the spin orientations.  $\rm Beta(\cdot)$ denotes the Beta
distribution, $\mathcal{N}_{[-1,1]}$ denotes a truncated Gaussian distribution
in the range $[-1,1]$, and $\alpha_{\chi}$ and $\beta_{\chi}$ are the parameters
of the Beta distribution.  More description and prior choices of the
hyperparameters are shown in Tab.~\ref{table:5}. We adopt the prior ranges following 
the \texttt{GWPopulation} package\footnote{\url{https://colmtalbot.github.io/gwpopulation/examples}} 
as a test case to validate the effectiveness of our method. 
More diverse prior choices can be found in previous studies~\cite{LIGOScientific:2020kqk,KAGRA:2021duu}.

%---------------------------------------------------------------------
\begin{table*}[]
    \renewcommand\arraystretch{1.5}
    \caption{Definitions and prior distributions for the hyperparameters used in
    our population inference with the \texttt{GWpopulation}
    package~\cite{Talbot:2024yqw}. The notation $\mathcal{U}(\theta; a,b)$
    represents a uniform prior for the parameter $\theta$ over the interval $[a,
    b]$.}
    \begin{tabularx}{\linewidth}{p{4cm}p{11cm}p{3cm}} 
    \hline\hline
    Notation & Parameter description & Prior \\
    \hline
    $\alpha$        & Negative power law exponent for the more massive black hole        & $\mathcal{U}(\alpha;-2,4)$ \\
    $\beta$         & Power law exponent of the mass ratio distribution & $\mathcal{U}(\beta;-4,12)$ \\
    $m_{\rm min}$       & Minimum black hole mass (${\rm M}_\odot$)                                       & $\mathcal{U}(m_{\rm min};2,2.5)$ \\
    $m_{\rm max}$       & Maximum black hole mass (${\rm M}_\odot$)                                        & $\mathcal{U}(m_{\rm max};80,100)$ \\
    $\lambda_{m}$   & Fraction of black holes in the Gaussian component              & $\mathcal{U}(\lambda_{m};0,1)$ \\
    $\mu_{m}$       & Mean of the Gaussian component    (${\rm M}_\odot$)                              & $\mathcal{U}(\mu_{m};10,50)$ \\
    $\sigma_{m}$    & Standard deviation of the Gaussian component  (${\rm M}_\odot$)                  & $\mathcal{U}(\sigma_{m};1,10)$ \\
    $\alpha_{\chi}$ & Parameters of Beta distribution for both black holes           & $\mathcal{U}(\alpha_{\chi};1,6)$ \\
    $\beta_{\chi}$  & Parameters of Beta distribution for both black holes           & $\mathcal{U}(\beta_{\chi};1,6)$ \\
    $\xi$           & Fraction of black holes in preferentially aligned component    & $\mathcal{U}(\xi;0,1)$ \\
    $\sigma$        & Width of preferentially aligned component                      & $\mathcal{U}(\sigma;0.3,4)$ \\
    \hline
    \end{tabularx}
  \label{table:5}
\end{table*}
%---------------------------------------------------------------------

\bibliography{reflibrary}

\end{document}